\documentclass[12pt]{article}
\usepackage{amssymb}
\def\Box{\square}

\newcommand{\rand}[1]{}
\newcommand{\labell}[1]{\label{#1}\rand{#1}}
\newcommand{\citel}[1]{\cite{#1}\rand{#1}}
\newcommand{\beql}[1]{\rand{#1}\begin{equation}\label{#1}}

\newcommand{\SECTION}[1]{\section{#1}\setcounter{equation}{0}}

\def\eeq{\end{equation}}
\def\ba{\begin{array}}
\def\ea{\end{array}}

\def\wt{\widetilde}
\def\wh{\widehat}

\def\R{{\mathbb R}} 
\def\N{{\mathbb N}} 
\def\C{{\mathbb C}}

\def\OO{{\cal O}} 
\def\PPP{{\bf Proof:} }
\def\QQQ{\mbox{}\hfill$\Box$\vspace{6mm}} 
\def\pl{\partial}
\def\ti{{\times}}
\def\fr{\frac}
\def\veps{\varepsilon}
\newcommand{\reff}[1]{{\rm (\ref{#1})}}
\newcommand{\set}[2]{ \{\, #1 \, | \, #2 \,\} }

\newtheorem{theorem}{Theorem}[section]
\newtheorem{lemma}[theorem]{Lemma}
\newtheorem{proposition}[theorem]{Proposition}
\newtheorem{corollary}[theorem]{Corollary}
\newcommand{\bsatz}{\begin{theorem} \hspace{-1mm}\sl\nopagebreak \\ }
\newcommand{\esatz}{ \end{theorem} \rm }
\newcommand{\blem}{\begin{lemma} \hspace{-1mm}\sl\nopagebreak \\ }
\newcommand{\elem}{\end{lemma} \rm }
\newcommand{\bprop}{\begin{proposition} \hspace{-1mm}\sl\nopagebreak \\ }
\newcommand{\eprop}{\end{proposition} \rm }
\newcommand{\bcoro}{\begin{corollary} \hspace{-1mm} \sl\nopagebreak \\ }
\newcommand{\ecoro}{\end{corollary} \rm }

\newcommand{\brem}{\smallskip \noindent {\bf Remark.}\ }
\newcommand{\brems}{\smallskip \noindent {\bf Remarks.}\ }
\newcommand{\erem}{\smallskip \newline }

\def\Re{\mathrm{Re}\,}

\def\n{\wt{\eta}}
\def\H{{\mathrm H}}      
\def\L{{\mathrm L}}      
\def\S{{\mathcal S}}     
\def\W{{\mathrm W}}      
\def\B{{\mathcal B}}     
\def\Tr{{\mathrm{Tr}}}   
\def\Det{{\mathrm{Det}}} 
\def\tp{{\mathrm t}}     
\def\CC{{\mathcal C}}    
\def\LL{{\mathbf L}}     
\def\KKK{{\mathbf K}}    
\def\LLL{{\mathcal L}}   
\def\ess{{\mathrm{ess}}} 
\def\phi{\varphi}

\newcommand{\e}{{\mathrm e}}    
\renewcommand{\d}{{\mathrm d}}  
\renewcommand{\i}{{\mathrm i}}  
\def\wtN{\wt{N}}

\def\INTER{{(-\fr1{\sqrt3},\fr1{\sqrt3})}}
\def\epsilon{\varepsilon}
\def\lu{{}_{\mathrm{lu}}}
\def\Hloc#1{{{\mathrm H}\lu^{#1}(\R)}}
\def\Efour{G_2}          
\def\Efive{G_3}          
\def\Esix{H_2}           

\begin{document}
\title{Diffusive Mixing of Stable States in the \\
Ginzburg--Landau Equation}
\author{Thierry Gallay\thanks{Analyse Num\'erique et EDP, Universit\'e de 
Paris XI, {\tt Thierry.Gallay@math.u-psud.fr}}
\ and Alexander Mielke\thanks{Institut f\"ur Angewandte
Mathematik, Universit\"at Hannover, {\tt mielke@ifam.uni-hannover.de}} }
\date{January 22, 1998}
\maketitle

\begin{abstract} 
The Ginzburg-Landau equation $\pl_tu=\pl_x^2u+u-|u|^2u$ on the real line 
has spatially periodic steady states of the the form $U_{\eta,\beta}(x)=
\sqrt{1{-}\eta^2}\,\e^{\i (\eta x+\beta)}$, with $|\eta| \leq 1$ and $\beta 
\in \R$. For $\eta_+,\eta_-{\in}\INTER$, $\beta_+,\beta_-{\in}\R$, we 
construct solutions which converge for all $t>0$ to the limiting pattern 
$U_{\eta_\pm,\beta_\pm}$ as $x\to \pm \infty$. These solutions are stable 
with respect to sufficiently small $\H^2$ perturbations, and behave 
asymptotically in time like $(1-\n(x/\sqrt t)^2)^{1/2}\,\exp(\i\sqrt t\,
\wtN(x/ \sqrt t\,))$, where $\wt N'=\n \in \CC^\infty(\R)$ is uniquely 
determined by the boundary conditions $\n(\pm\infty) = \eta_\pm$. 
This extends a previous result of \citel{BricKup} by removing the assumption 
that $\eta_\pm$ should be close to zero. The existence of the limiting
profile $\n$ is obtained as an application of the theory of monotone 
operators, and the long-time behavior of our solutions is controlled 
by rewriting the system in scaling variables and using energy 
estimates involving an exponentially growing damping term. 
\end{abstract}

\vfill\eject

\SECTION{Introduction}

The dynamics of parabolic partial differential equations on unbounded spatial 
domains exhibit various interesting phenomena which are not present on 
bounded domains, since the semigroups associated to such extended systems
are not compact. 
An important example is the development of so-called ``side-band 
instabilities'' for spatially periodic patterns bifurcating from a homogeneous
steady state. Although this problem has been investigated at a formal level
more than 30 years ago \citel{Eckhaus}, only recently a rigorous approach 
has been proposed and applied to various model problems, see \citel{CoEckm90},
\citel{Mielke} and the references therein. In particular, nonlinear stability 
results have been obtained for the spatially periodic solutions of the 
one-dimensional Ginzburg-Landau \citel{CoEckmEp92} and Swift-Hohenberg 
\citel{SchnCMP} equations, as well as for the Taylor vortices in a cylindrical
geometry \citel{SchnHABIL}. These periodic patterns appear to be stable with
respect to sufficiently localized perturbations, and the convergence of the 
perturbed solutions towards the equilibrium state is only diffusive (typically 
$\OO(t^{-1/2})$) as $t \to +\infty$. In higher dimensions, we quote the 
recent results by \citel{Mi-SHE} and \citel{Uecker} which prove the linear 
and nonlinear stability of the roll solutions for the two-dimensional 
Swift-Hohenberg equation. 

Following up the work of Collet-Eckmann \citel{CoEckm92} and 
Bricmont-Kupiainen \citel{BricKup}, we study here one particular aspect
of the diffusive stability theory, namely the mixing of two spatially periodic
steady states. Consider a partial differential equation on the real line
having a continuous family of spatially periodic time-independent solutions, 
some of them being (diffusively) stable. For this system, we prescribe initial 
data which coincide with one stable pattern as $x \to -\infty$ and with 
another one as $x \to +\infty$. The question is to determine the long-time 
behavior of the solutions arising from such initial data. 

As in \citel{CoEckm92}, \citel{BricKup}, we investigate this problem for 
the Ginzburg-Landau equation with real coefficients
\beql{e2.1}
 \pl_t u=\pl_x^2u +u-|u|^2u\,, \quad t\geq 0\,,\quad x\in \R\,,
\eeq
where $u(t,x)\in \C$. This system has a family of periodic stationary states 
given by
\[
 U_{\eta,\beta}(x)=\sqrt{1-\eta^2}\:\e^{\i (\eta x+\beta)},
\]
where $|\eta|\leq 1$ and $\beta\in \R$. These solutions are linearly stable
if and only if $\eta^2 \leq 1/3$ \citel{Eckhaus}, and nonlinearly stable
with respect to $\H^1$ perturbations for all $\eta^2 < 1/3$ 
\citel{CoEckmEp92}, \citel{Gallay}. Thus, we shall study solutions of 
\reff{e2.1} satisfying 
\beql{e2.2}
 u(t,x) - U_{\eta_\pm,\beta_\pm}(x) \to 0 \quad\mbox{as }x\to \pm\infty\,,
\eeq
with $\eta_+,\eta_- \in \INTER$ and $\beta_+,\beta_- \in \R$. 

If $\eta_+=\eta_-$ and $\beta_+=\beta_-$, the problem reduces to the diffusive
stability of the steady state $U_{\eta_-,\beta_-}$. In the sequel, we shall
concentrate on the most interesting case $\eta_+ \neq \eta_-$, and make only
a few comments on the intermediate situation where $\eta_+=\eta_-$ and 
$\beta_+ \neq \beta_-$. Thus, we assume that $\eta_+ \neq \eta_-$ and, 
without loss of generality, that $\beta_+ = \beta_- = 0$, see 
\reff{e2.choice} below. If both $\eta_+,\eta_-$ are close to zero, it is 
shown in \citel{BricKup} that, under technical assumptions on 
the initial data, the solutions of \reff{e2.1} satisfying \reff{e2.2}
converge locally as $t \to +\infty$ to the circle of steady states 
${\mathbf S}_* = \set{U_{\eta_*,\beta} }{ \beta \in [0,2\pi] }$, where 
$\eta_* \in \INTER$ is uniquely determined by the boundary values $\eta_+,
\eta_-$. The main purpose of this paper is to generalize this result to 
{\sl all} values of $\eta_+,\eta_- \in \INTER$. 

For our analysis to work, it will be necessary to make assumptions on the
behavior 
of the solutions not only at infinity, but also in the transition region 
connecting the two limiting patterns. In particular, we shall assume that the
solutions stay locally close to the family of steady states 
$\set{U_{\eta,\beta} }{ |\eta| \leq 1\,,\,\beta \in \R }$. Writing 
$u=r\,\e^{\i\phi}$, this implies that the amplitude $r$ is slaved to
the phase derivative $\wh\eta=\pl_x \phi$ in such a way that 
$r^2+\wh\eta^2\approx 1$ for all times. In fact, we shall construct a class
of solutions for which $r^2+\wh\eta^2 = 1 + \OO(1/t)$ as $t \to +\infty$. 
The behavior of these solutions will be described in a good approximation
by the {\sl phase diffusion equation}
\beql{e2.phase}\textstyle
 \pl_t\phi= \pl_x[A(\pl_x\phi)]\, ,\quad
 \mbox{where }A(\eta)=\int_0^\eta a(s)\,\d s\,, \quad a(s) = 
 \fr{1-3s^2}{1-s^2}\,. 
\eeq
This equation is obtained from \reff{e2.1} by rewriting it in polar 
coordinates $(\phi,r)$, and then substituting $r=\sqrt{1{-}(\pl_x\phi)^2}$ 
into the equation for $\phi$. 

{}From the results in \citel{BricKup}, it is apparent that the long-time
dynamics of Eq.\reff{e2.phase} is most naturally described in the scaling
coordinates $\tau =\log t$, $\xi=x/\sqrt t$. Thus, defining $\eta(\tau,\xi)
=\pl_x\phi(\e^\tau,\e^{\tau/2}\xi)$, the equation \reff{e2.phase} is 
transformed into
\beql{e1.4}
\dot \eta =[A(\eta)]''+\fr\xi2\eta',\quad \mbox{where }\dot{(\ )}=\pl_\tau(\ )
 \mbox{ and }(\ )'=\pl_\xi(\ ),
\eeq
and the boundary conditions \reff{e2.2} become
\beql{e1.bdc}
\lim_{\xi\to -\infty}\eta(\tau,\xi)=\eta_-\, ,\quad
\lim_{\xi\to +\infty}\eta(\tau,\xi)=\eta_+\,.
\eeq

In Section \ref{sect.phase} below we show that, for all
$\eta_+,\eta_-{\in}\INTER$,  
Eq.\reff{e1.4} has a unique steady state $\n(\xi)$ satisfying the boundary 
conditions \reff{e1.bdc}. This profile is a monotone function of $\xi$ and 
approaches its limits $\eta_\pm$ faster than any exponential. If $|\eta_\pm|
\ll 1$, these results were already obtained in \citel{BricKup} using a 
perturbation argument based on the implicit function theorem. In this 
case, one has the approximate expression
\[
  \n(\xi) \approx \eta_- + \frac{\eta_+ - \eta_-}{\sqrt{4\pi}} \,
  \int_{-\infty}^\xi \e^{-z^2/4}\,\d z ~.
\]
Our construction is nonperturbative and relies on the monotonicity properties
of the differential operator in the right-hand side of \reff{e1.4}. At 
the same time, our monotonicity argument shows that $\n$ is a stable 
equilibrium point of \reff{e1.4}, in the sense that any solution 
$\eta(\tau,\xi)$ which is sufficiently close to $\n$ in the $\H^2$ norm 
converges exponentially to $\n$ as $\tau \to +\infty$. Returning to the
original variables, this implies that $\phi(t,x) \approx \sqrt{t} 
\wtN(x/\sqrt t)$ as $t \to +\infty$, where $\wtN(\xi)=\eta_-\xi + 
\int_{-\infty}^\xi [\n(s)-\eta_-]\,\d s$. 

We are now able to state our main result in a precise way. Let $\eta_+,
\eta_- \in \INTER$, $\eta_+ \neq \eta_-$, and assume that $\beta_+ = \beta_-
= 0$. For $t > 0$, we define the time-dependent profile 
\beql{e1.profile}
 \wt U(t,x)=\sqrt{1-\n(x/\sqrt t\,)^2}\, \e^{\i \sqrt t\,\wtN(x/ \sqrt t\,)}\,,
\eeq
where $\n$ and $\wtN$ are constructed above. We consider initial data 
$u_0 \in \H^2_{\mathrm{loc}}(\R)$ such that $\|u_0(\cdot)-\wt 
U(t_0,\cdot)\|_{\H^2} \leq \epsilon$ for $\epsilon > 0$ sufficiently small
and $t_0 > 0$ sufficiently large. Then Theorem~\ref{thm4.2} below states that 
the solution $u(t,x)$ of \reff{e2.1} with initial condition $u(0,\cdot) = 
u_0$ satisfies
\beql{e1.5}
 \|u(t,\cdot)-\wt U(t_0{+}t,\cdot)\|_{\L^\infty} \,=\, \OO(t^{-\nu/4})
 \quad \mbox{as } t \to +\infty\,,
\eeq
for any $\nu \in (0,1)$. In particular, setting $\eta_* = \n(0)$ and
$\phi_* = \wtN(0)\neq 0$, we have for all $x_0 > 0$:
\beql{e1.6}
 \sup_{|x| \leq x_0} |u(t,x)-U_{\eta_*,\phi_*\sqrt{t}}(x)| \,=\, 
 \OO(t^{-\nu/4}) \quad \mbox{as } t \to +\infty\,,
\eeq
see Corollary~\ref{cor4.3}. 

This result shows that the solution $u(t,\cdot)$ converges uniformly on 
compact sets to the circle of steady states ${\mathbf S}_* = 
\set{U_{\eta_*,\beta} }{ \beta \in [0,2\pi] }$. Following \citel{MiSchn}, 
we can also study the convergence in the weighted norm $\H^1_w$ and the 
uniformly local norm $\H\lu^1$ (see \reff{e4.weight} below for the
definitions). We find 
\[
 \lim_{t\to\infty}{\mathrm{dist}}_{\H^1_w}(u(t,\cdot),{\mathbf S}_*)= 0\,,
 \quad \mbox{and }
 \liminf_{t\to\infty}{\mathrm{dist}}_{\H\lu^1}(u(t,\cdot),{\mathbf S}_*) 
>0\, .
\]
Indeed, in the $\H^1_w$ topology the solution $u(t,\cdot)$ approaches 
${\mathbf S}_*$ sliding along the circle with nonzero speed $\phi_*t^{-1/2}$, 
hence the $\omega$-limit set of $u(t,\cdot)$ is equal to ${\mathbf S}_*$. 
In the $\H\lu^1$ topology the $\omega$-limit set of $u(t,\cdot)$ is empty. 

The main ingredient in the proof of our convergence results is the use of 
the nonlinear change of variables defined by $u=r\,\e^{\i\phi}$ and
\beql{e1.newvar}
 \psi(\tau,\xi)=\e^{-\tau/2}\phi(\e^\tau,\e^{\tau/2}\xi)-\wt N(\xi)\,, \quad
 s(\tau,\xi)=\log\left(\fr{r(\e^\tau,\e^{\tau/2}\xi)^2}{1-[\pl_x\phi
 (\e^\tau,\e^{\tau/2}\xi)]^2}\right)\,.
\eeq
Remark that $s = 0$ is equivalent to $r^2 + (\pl_x\phi)^2 = 1$, hence 
$s$ measures the slaving of the amplitude to the phase derivative.
Note also that $s$ has the same regularity as $\psi' = \pl_\xi\psi$. 
The new functions $(\psi,s)$ are solutions of a quasilinear nonautonomous 
parabolic system given by \reff{e2.12} below. In Section~\ref{sect.local}, 
we show that this system is locally well-posed in the space $\H^2(\R) \ti 
\H^1(\R)$. In Section~\ref{sect.energy}, we prove using various energy 
estimates that the small solutions $(\psi,s)$ of \reff{e2.12} satisfy
\beql{e1.decay}
  \|\psi(\tau)\|_{\H^2(\R)}=\OO(\e^{-\gamma \tau})\,,\quad
  \| s(\tau) \|_{\H^1(\R)}=\OO(\e^{-\tau})\,,\quad \tau \to +\infty\,,
\eeq
for all $\gamma\in(0,\fr34)$, see Theorem~\ref{thm4.1}. Returning to the
original variables, this implies the convergence result \reff{e1.5}. 

We conclude this section with a few comments on possible generalizations. 
Without any additional difficulty, we can extend our results to cover the
equation
\[
  \pl_t u = \pl_x^2 u + h(|u|^2) u\,,
\]
where $h:[0,\infty)\to \R$ is a smooth, decreasing function such that 
$h(0) > 0$ and $h(s) < 0$ for sufficiently large $s$. This equation has
stationary sates $U_{\eta,\beta} = H(\eta^2)^{1/2}\e^{\i (\eta x{+}\beta)}$,
where $|\eta| \leq h(0)$ and $H = h^{-1}$. The Eckhaus stability criterion
takes the form $H(\eta^2) + 2\eta^2 H'(\eta^2) > 0$, and thus is satisfied
for all $\eta$ in a symmetric interval $(-\eta_0,\eta_0)$. The phase 
diffusion equation \reff{e2.phase} becomes
\[
  \pl_t \phi = \pl_x  \big[ A(\pl_x\phi)\big]\,,\quad\mbox{where }
  A(\eta)=\int_0^\eta [1+2s^2H'(s^2)/H(s^2)]\, \d s.
\]
For all $\eta_+,\eta_- \in (-\eta_0,\eta_0)$, the arguments of 
Section~\ref{sect.phase} show the existence and uniqueness of the steady
state $\n$ satisfying \reff{e1.4}, \reff{e1.bdc}, and the stability of
this profile is proved as in Sections~\ref{sect.local} and \ref{sect.energy}.

Finally, an interesting open problem is the diffusive mixing in  
parabolic equations without phase invariance, such as the Swift-Hohenberg
equation  
\beql{e1.SHE}
  \pl_t u = -(1+\pl_x^2)^2 u +\veps u- u^3\,, \quad \epsilon > 0\,.
\eeq
This equation also admits (if $\epsilon$ is sufficiently small) a 
continuous family of spatially periodic steady states, whose stability 
has been proved recently \citel{SchnCMP}, \citel{EckWayWit}. 
In addition, the solutions of \reff{e1.SHE} are well approximated 
for a long (but finite) interval of time by those of \reff{e2.1}, 
see \citel{CoEckm90'}, \citel{KirSchMi}. Therefore, a natural question 
is whether the results of this paper can be extended to the higher order
equation \reff{e1.SHE}. 

The rest of the paper is organized as follows. In Section~\ref{sect.reform}, 
we reformulate our problem in terms of the scaling variables $\tau,\xi$ 
and the new functions $\psi,s$. In Section~\ref{sect.phase}, we prove the
existence of the steady profile $\n$, and give a few general results on 
the dynamics of the phase diffusion equation \reff{e1.4}.  
Section~\ref{sect.results} contains the precise statements of our convergence
results in both the original and the rescaled variables. The local existence
of solutions for the $(\psi,s)$ system \reff{e2.12} is proved in  
Section~\ref{sect.local}, and Section~\ref{sect.energy} is devoted to 
the proof of our main result (Theorem~\ref{thm4.1}) using energy estimates.

\subsubsection*{Acknowledgments.}
The authors are grateful to G. Raugel and G. Schneider for helpful
discussions. This work was begun when A.M. visited the University of 
Paris~XI and Th.G. the University of Hannover. Both institutions are 
acknowledged for their hospitality. This research was partially supported 
by the DFG-Schwerpunktprogramm `Dynamische Systeme' under the grant 
Mi 459/2-2.


\SECTION{Reformulation of the problem}
\labell{sect.reform}

Let $\eta_+,\eta_- \in \INTER$, $\beta_+,\beta_- \in \R$, and let 
$u(t,x)$ be a solution of \reff{e2.1} satisfying $u(t,x) \neq 0$ for all
$t,x$. Introducing the polar coordinates $(\phi,r)$ via $u = r\e^{\i\phi}$,
we assume that the boundary conditions \reff{e2.2} hold, namely
\[
  \phi(t,x) - \eta_\pm x - \beta_\pm \to 0~, \quad r(t,x) \to 
  \sqrt{1-\eta_\pm^2}~, \quad\mbox{as } x \to \pm\infty~.
\]
We suppose from now on that $\eta_+ \neq \eta_-$, and refer to the end of 
this section for a comment on the simpler case $\eta_+ = \eta_-$. 
We first observe that, due to the translation and phase symmetries of the
Ginzburg-Landau equation \reff{e2.1}, there is no loss of generality in 
assuming $\beta_+ = \beta_- = 0$. Indeed, this amounts to replacing $u(t,x)$
with $u(t,x-x_0)\e^{-\i\phi_0}$, where
\beql{e2.choice}
  x_0 = \fr{\beta_+ - \beta_-}{\eta_+ - \eta_-}~, \quad
  \phi_0 = \fr{\eta_+\beta_- - \eta_-\beta_+}{\eta_+ - \eta_-}~.
\eeq
Assuming thus $\beta_\pm = 0$, we obtain for $(\phi,r)$ the system
\beql{e2.4}
  \pl_t\phi=\pl_x^2\phi+ 2\fr{\pl_x r }{r}\pl_x\phi~,\quad 
  \pl_t r=\pl_x^2 r+r(1-r^2-\big[\pl_x\phi\big]^2)~,
\eeq
together with the boundary conditions $\phi(t,x) - \eta_\pm x \to 0$, 
$r(t,x) \to \sqrt{1-\eta_\pm^2}$ as $x \to \pm\infty$. 
For later use, we also introduce 
the local wave-vector $\wh \eta(t,x) =\pl _x \phi(t,x)$, which satisfies
\[
  \pl_t\wh\eta=\pl_x^2\wh\eta+2\pl_x\left(\fr{\pl_x r}{r}\wh\eta\right)~,
  \quad \wh\eta(t,x) \to \eta_\pm \quad\mbox{as }x \to \pm\infty~.
\]

As explained in the introduction, we shall use the scaling coordinates
$\tau=\log t$, $\xi =x/\sqrt{t}$ to investigate the long-time dynamics
of the system \reff{e2.4}. Defining 
\[
  \eta(\tau,\xi)=\wh\eta(\e^\tau,\e^{\tau/2}\xi)~, \quad
  \rho(\tau,\xi)=r(\e^\tau,\e^{\tau/2}\xi)~,
\]
we obtain the rescaled system
\beql{e2.etarho}
 \dot\eta=\eta''+\fr\xi2\eta' + 2\left(\fr{\rho'\eta}{\rho}\right)'~, \quad
 \dot\rho=\rho''+\fr\xi2\rho'+\e^\tau\rho\left(1-\rho^2-\eta^2\right)~,
\eeq
where $\dot{(\ )}=\pl_\tau(\ )$ and $(\ )'=\pl_\xi(\ )$. This nonautonomous 
system has an asymptotic equilibrium point $(\n,\wt\rho)$ (as $\tau \to 
+\infty$) defined by $\wt\rho\,^2 = 1-\n\,^2$ and
\beql{e2.ODE}
  \big[A(\n)\big]'' + \fr\xi2\n' = 0~, \quad \mbox{where }
  A(\eta) = \int_0^\eta \fr{1-3s^2}{1-s^2}\,\d s~.
\eeq
In Theorem \ref{thm3.1} below, we show that the differential equation 
\reff{e2.ODE} for $\n$ has a unique solution satisfying 
$ \n(\xi) \to \eta_\pm$ as $\xi\to \pm \infty$. This solution is monotone in
$\xi$ and approaches its limits $\eta_\pm$ faster than any exponential. In
addition, we have
\beql{e2.step}
  \int_\R \big[\n(\xi)-\eta_\infty(\xi)\big]\,\d\xi = 0~, \quad 
  \mbox{where } \eta_\infty(\xi) = \cases{\eta_+ & if $\xi > 0$ \cr \eta_- & 
  if $\xi < 0$}~. 
\eeq

The main purpose of this paper is to prove that, for suitable initial data, 
the solution $(\eta,\rho)$ of \reff{e2.etarho} converges to $(\n,\wt\rho)$ 
as $\tau \to +\infty$. For the phase $\phi(t,x)$, this will imply the
long-time behavior
\beql{e2.phi}
  \phi(t,x) \,\approx\, \eta_-x + \sqrt{t}\int_{-\infty}^{x/\sqrt t}  
  \big[\n(\xi)-\eta_-\big]\,\d\xi \, \equiv\, \eta_+x - \sqrt{t}
  \int_{x/\sqrt t}^{+\infty}\big[\n(\xi)-\eta_+\big]\,\d\xi~, 
\eeq
since $\pl_x \phi(t,x) = \wh\eta(t,x) \approx \n(x/\sqrt{t})$ and 
$\phi(t,x) - \eta_\pm x \to 0$ as $x \to \pm\infty$ (note that the two 
expressions in the right-hand side of \reff{e2.phi} are identical due to 
\reff{e2.step}). This  motivates the following ansatz for $\phi(t,x)$. Let
\beql{e2.tildeN}
  \wtN(\xi) \,=\,\eta_-\xi +\int_{-\infty}^\xi \big[\n(s)-\eta_- \big]\,\d s
  \,\equiv\, \eta_+\xi -\int_\xi^{+\infty} \big[\n(s)-\eta_+ \big]\,\d s~.
\eeq
We define
\beql{e2.5}
  \psi(\tau,\xi)=\e^{-\tau/2}\phi(\e^\tau,\e^{\tau/2}\xi) -\wtN(\xi),\quad
  \rho(\tau,\xi)=r(\e^\tau,\e^{\tau/2} \xi), 
\eeq
or equivalently
\beql{e2.5a}
  \phi(t,x)=\sqrt t\,\big[\wtN(x/\sqrt{t})+ \psi(\log t,x/\sqrt t\,)\big], 
  \quad r(t,x)=\rho(\log t,x/\sqrt t).
\eeq
Then $\psi(\tau,\xi) \to 0$ as $\xi \to \pm\infty$, and \reff{e2.4} is 
transformed into
\beql{e2.6}\ba{rcl}
 \dot\psi&=&(\wtN+\psi)''+\fr\xi2(\wtN+\psi)'-\fr12(\wtN+\psi)+
             2\fr{\rho'}{\rho}\eta~, \\[1.5mm]
 \dot\rho&=&\rho''+\fr\xi2\rho'+\e^\tau\rho\big(1-\rho^2-\eta^2\big)~,
\ea\eeq
where $\eta = (\wtN+\psi)' = \n + \psi'$. 

By construction, we expect that the solution $(\psi,\rho)$ of \reff{e2.6}
converges to $(0,\wt\rho)$ as $\tau \to +\infty$, where $\wt\rho\,^2 = 1-
\n\,^2$. Assuming this to hold, we deduce from the second equation in 
\reff{e2.6} that $\rho(\tau,\xi)^2 = 1 - \eta(\tau,\xi)^2+\OO(\e^{-\tau})$. 
This is the so-called ``slaving'' of the amplitude $\rho$ to the phase 
derivative $\eta$. To take this behavior into account, we parametrize the 
amplitude $\rho$ with the new variable $s$ defined by
\beql{e2.9}
  \rho = \sqrt{1-\eta^2}\, \e^{s/2} \equiv \sqrt{1-(\n+\psi')^2}\,\e^{s/2}\,.
\eeq
After some elementary calculations, we obtain the following equations for
$\psi,s$
\beql{e2.12}\ba{rcl}
  \dot\psi&=&\big[A(\n+\psi')-A(\n)\big]'+\fr\xi2\psi'-
  \fr12\psi+\eta s'~, \\[1mm]
\dot s&=&a_1(\eta)s''+a_2(\eta)\eta''+\fr\xi2 s' -2\e^{\tau}(1{-}\eta^2)
   (\e^s{-}1)+\fr12s'^2+a_3(\eta)\eta'^2~,
\ea\eeq
where $\eta = \n + \psi'$. The coefficients  $a = A'$ and
$a_1,a_2,a_3$ are given by
\beql{e2.aaa}
  a(\eta)=\fr{1{-}3\eta^2}{1{-}\eta^2}~,\quad
  a_1(\eta)=\fr{1{+}\eta^2}{1{-}\eta^2}~,\quad
  a_2(\eta)=\fr{-4\eta^3}{(1{-}\eta^2)^2}~,\quad
  a_3(\eta)=\fr{-2{-}6\eta^2}{(1{-}\eta^2)^3}~.
\eeq

This is the final form of the problem which is the basis of our analysis in
Sections \ref{sect.local} and \ref{sect.energy}. In particular, our main
result (Theorem~\ref{thm4.1}) will show that sufficiently small solutions 
$(\psi,s)$ of \reff{e2.12} satisfy the decay estimate \reff{e1.decay}. 
Remark that there is now an imbalance in the number of spatial derivatives 
of $\psi$ and $s$ in \reff{e2.12}, since the $s$--equation contains $\eta''
= \n'' + \psi'''$, whereas only $s'$ appears in the $\psi$--equation. 
This derives from the fact that $\rho$ is slaved to $\eta=\n+\psi'$. 
Due to this imbalance, we also have to consider the equation for $\delta=
\psi'=\eta-\n$ which reads
\beql{e2.13}\textstyle
  \dot \delta= \big[A(\n{+}\delta) -A(\n)\big]''+\fr\xi2\delta'+
  ((\n{+}\delta) s')'~.
\eeq

\brem
When $\eta_+ = \eta_-$, the discussion above remains valid, except for an
important difference : if $\beta_+ \neq \beta_-$, it is no longer possible 
to reduce the problem to $\beta_\pm = 0$ using the symmetries of the 
Ginzburg-Landau equation. As a consequence, the function $\psi(\tau,\xi)$
defined by \reff{e2.5} has non trivial boundary values $\psi(\tau,\pm\infty)
= \beta_\pm$, and $\int \delta(\tau,\xi)\d\xi = \beta_+ - \beta_- \neq 0$. 
It follows that $\delta(\tau,\cdot)$ converges to zero like $e^{-\tau/2}$
as $\tau \to +\infty$, and not faster as in the previous case. Indeed, since
$\n(\xi) = \eta_+ = \eta_-$ is now constant, the linearized equation
\reff{e2.13} becomes
\[\textstyle
  \dot \delta = \wt{a}\delta'' + \fr\xi2\delta' + \n s''~, \quad
  \mbox{where }\wt{a} = a(\n)~.
\]
As is well-known, the largest eigenvalue of the operator $\wt{a}\pl_\xi^2 + 
\fr\xi2\pl_\xi$ is $-1/2$, with eigenfunction $\omega(\xi) = (4\pi\wt{a})^{-1/2}
\exp(-\xi^2/4\wt{a})$ and adjoint eigenfunction $1$. Since $s(\tau,\cdot) = 
\OO(e^{-\tau})$ due to the slaving, we conclude that $\delta(\tau,\xi) \approx
e^{-\tau/2}m(\xi)$ as $\tau \to +\infty$~, where $m(\xi) = (\beta_+ - \beta_-)
\omega(\xi)$. Therefore, a natural ansatz for $\psi$ is $\psi(\tau,\xi) = 
e^{-\tau/2}M(\xi) + \chi(\tau,\xi)$, where
\[
   M(\xi) = \beta_- + \fr{\beta_+ - \beta_-}{(4\pi\wt{a})^{1/2}} 
   \int_{-\infty}^\xi \e^{-y^2/(4\wt{a})}\,\d y~, \quad M'(\xi) = m(\xi)~.
\]
With this definition, the equations for $\chi,s$ are
\beql{e2.14}\ba{rcl}
  \dot\chi&=&\big[A(\n+e^{-\tau/2}m + \chi')-A(\n+e^{-\tau/2}m)
   \big]'+\fr\xi2\chi'- \fr12\chi + \eta s' + \Lambda~, \\[1mm]
\dot s&=&a_1(\eta)s''+a_2(\eta)\eta''+\fr\xi2 s' -2\e^{\tau}(1{-}\eta^2)
   (\e^s{-}1)+\fr12s'^2+a_3(\eta)\eta'^2~,
\ea\eeq 
where $\eta = \n+e^{-\tau/2}m + \chi'$ and $\Lambda(\tau,\xi) = 
e^{-\tau/2}\big[a(\n+e^{-\tau/2}m)-a(\n)\big]m' = \OO(e^{-\tau})$. 
Then, proceeding as in Section~\ref{sect.energy}, one can show that 
$\|\chi(\tau,\cdot)\|_{\H^2} = \OO(\e^{-\gamma\tau})$ and 
$\|s(\tau,\cdot)\|_{\H^1} = \OO(\e^{-\tau})$ as $\tau \to +\infty$, for all
$\gamma \in (0,3/4)$. 
\erem

\brem
An important features of the Ginzburg-Landau equation are its symmetry
properties which are essential 
to our analysis in several respects. Eq.\ \reff{e2.1} is invariant under
the phase rotations $P_\alpha$ and the translations in time $T_{z}$ and space 
$S_{y}$ defined by
\beql{e2.3}\ba{rll}
 T_{z}:&(t,x,u)\mapsto (t+z,x,u), &z \in \R;\\
 S_{y}:&(t,x,u)\mapsto (t,x+y,u), &y \in \R;\\
P_{\alpha}:&(t,x,u)\mapsto (t,x,\e^{\i \alpha}u), &\alpha \in \R.
\ea\eeq
For the transformed system \reff{e2.etarho}, the rotation $P_\alpha$ is 
trivial, and the translations $T_z$, $S_y$ become
\beql{e2.7}\ba{rl}
T_{z}:&(\tau,\xi,\eta,\rho)\mapsto (\tau{+}\log(1+\e^{-\tau}z),
           \xi/\sqrt{1+\e^{-\tau}z},\eta,\rho);\\
S_{y}:&(\tau,\xi,\eta,\rho)\mapsto (\tau,\xi+\e^{-\tau/2}y,\eta,\rho).\\
\ea\eeq
The transformation rules for $\psi(\tau,\xi)$ and $\delta(\tau,\xi)$ are 
more complicated and will not be used in the sequel. 
\erem


\SECTION{The phase diffusion equation}
\labell{sect.phase}

In this section, we first prove the existence of a unique solution $\n(\xi)$ 
of \reff{e2.ODE} satisfying the boundary conditions $\n(\xi) \to \eta_\pm$ 
as $\xi \to \pm\infty$. We next study the structure and dynamics of the 
{\sl phase diffusion equation} (or {\sl slaved equation})
\beql{e3.1}
 \dot\psi=\big[A(\n+\psi')\big]'+\fr\xi2(\n+\psi')-\fr12(\wtN+\psi),
\eeq
which is obtained by setting $s=0$ in the $\psi$--equation of \reff{e2.12}. 
Since $s(\tau,\cdot)$ decays to zero like $e^{-\tau}$ as $\tau \to +\infty$ 
due to the slaving, it is reasonable to expect that the long-time dynamics of 
the full system \reff{e2.12} is well approximated by \reff{e3.1}, even though 
$s \equiv 0$ does not define an invariant submanifold. In addition, some
structures of the full problem can be understood better on this simpler model. 
We begin with the existence of the steady state:

\bsatz
(a)\labell{thm3.1}
For each $\eta_+,\eta_- \in \INTER $ there exists a unique
$\n\in \CC^2(\R,\INTER)$ solving
\beql{e3.4}
\big[ A(\n)\big]''+\fr\xi2 \n'=0 \mbox{ on }\R,\quad \mbox{and}\quad
\n(\xi)\to \eta_\pm\mbox{ as }\xi\to\pm\infty~,
\eeq
where $A(\eta)$ is defined in \reff{e2.ODE}. \\
(b) The solution $\n$ is constant if $\eta_+=\eta_-$ and strictly
monotone otherwise. In addition, for each $C>0$ we have
$\sup\set{\e^{C|\xi|}|\n'(\xi)|}{\xi\in\R} <\infty$.
\\
(c) The function $\n$ satisfies
\[
\int_\R\big[\n(\xi){-}\eta_\infty(\xi)\big]\,\d\xi = 0 \quad 
\mbox{and} \quad \int_\R\xi\big[\n(\xi){-}\eta_\infty(\xi)\big]\,\d\xi = 
A(\eta_-){-}A(\eta_+)~, 
\]
where $\eta_\infty$ is defined in \reff{e2.step}. Setting $\phi_*=
\int_{-\infty}^0\big[\n(\xi)-\eta_-\big]\,\d\xi = \int_0^{+\infty}
\big[\eta_+ - \n(\xi)\big]\,\d\xi$, we have ${\mathrm{sign}}(\phi_*)=
{\mathrm{sign}}(\eta_+{-}\eta_-)$, and hence $\phi_*\neq 0$ if $\eta_+\neq 
\eta_-$.
\esatz
\PPP
Let $\eta_+,\eta_- \in \INTER$. The existence of the steady profile 
$\n$ will be obtained using the theory of (nonlinear) monotone operators. 
First, choosing $\veps > 0$ sufficiently small, we modify $a(\eta)=
(1-3\eta^2)/(1-\eta^2)$ outside the interval containing $\eta_+, \eta_-$ 
in such a way that the resulting function (still denoted by $a$) is 
$\CC^\infty$ and satisfies $1\geq a(\eta) \geq \veps>0$ for all  
$\eta\in\R$. Accordingly, the modified primitive $A(\eta) = \int_0^\eta a(y)
\,\d y$ verifies $(\eta_1-\eta_2)(A(\eta_1)-A(\eta_2))\geq \veps
(\eta_1-\eta_2)^2$ for all $\eta_1,\eta_2 \in \R$. Next, we choose a smooth, 
monotone function $n : \R \to \R$ such that
\[
|n'(\xi)| \le C \e^{-|\xi|}~, \quad |n(\xi)-\eta_\infty(\xi)| \le 
C\e^{-|\xi|}~, \quad \int_\R \big[n(\xi)-\eta_\infty(\xi)\big]\,\d\xi = 0~,
\]
for some $C > 0$, where $\eta_\infty$ is given by \reff{e2.step}. In analogy 
with \reff{e2.tildeN}, we set
\[
  N(\xi)\, = \,\eta_-\xi +\int_{-\infty}^\xi \big[n(s)-\eta_- \big]\,\d s
  \, \equiv\, \eta_+\xi -\int_\xi^{+\infty} \big[n(s)-\eta_+ \big]\,\d s~,
\]
and we observe that $|N(\xi)-\xi n(\xi)| \le C\e^{-|\xi|}$. Finally, 
we define the operators $B_1:\H^1(\R)\to \H^{-1}(\R)$ and $B_2:D(B)\subset 
\H^1(\R) \to \H^{-1}(\R)$ by
\beql{e3.operator}\ba{rcl}
 D(B) &=& \set{\psi \in \H^1(\R) }{ \xi \psi'\in \H^{-1}(\R)}~,\\
 B_1(\psi) &=& -\big[ A(n{+}\psi')\big]' +\fr12\psi~, \\
 B_2\psi &=& -\fr\xi2 \psi'~.
\ea
\eeq
The program is to show that the nonlinear equation 
\beql{e3.monot}
  B_1(\psi)+B_2\psi = \fr12 [\xi n(\xi)- N(\xi)] \in \H^{-1}(\R)
\eeq 
has a unique solution $\wt\psi\in D(B) \subset \H^1(\R)$. Then, 
$\n=n+\wt\psi'$ will be the desired solution of \reff{e3.4}. 

To prove the existence and uniqueness of the solution of \reff{e3.monot}, 
we first observe that $B_1, B_2$ are coercive (or strongly monotone) 
operators, with 
\[\ba{rcl}
\langle B_1(\psi_1){-}B_1(\psi_2),\psi_1{-}\psi_2\rangle_{\H^{-1}\ti \H^1} 
 &\geq& \veps \|\psi'_1{-}\psi'_2\|^2_{\L^2}+\fr12 
 \|\psi_1{-}\psi_2\|^2_{\L^2}~\forall \psi_1,\psi_2 \in \H^1~, \\
\langle B_2(\psi_1{-}\psi_2),\psi_1{-}\psi_2\rangle_{\H^{-1}\ti \H^1} 
 &=& \fr14 \|\psi_1{-}\psi_2\|^2_{\L^2}~\forall \psi_1,\psi_2 \in D(B)~.
\ea
\]
Clearly the nonlinear operator $B_1$ is maximal monotone as it is continuous 
(from $\H^1$ into its dual $\H^{-1}$) and monotone, see \citel{ZeidlerII} 
Proposition~32.7. On the other hand, the linear operator $B_2$ is densely
defined, and a direct calculation shows that the adjoint operator $B_2^*$
satisfies $D(B_2^*) = D(B)$ and $B_2^* = -B_2 + \fr12$. Thus $B_2$ is closed, 
$B_2^*$ is monotone, hence $B_2$ is maximal monotone by \citel{ZeidlerII} 
Theorem~32.L. Since ${\mathrm{int}}D(B_1)\cap D(B_2) = \H^1(\R)\cap D(B)=
D(B) \neq \emptyset$, it follows from the ``sum theorem'' (see  
\citel{ZeidlerII} Theorem 32.I) that $B=B_1+B_2:D(B)
\to \H^{-1}(\R)$ is maximal monotone. As $B$ is also coercive, we conclude 
that $B$ is one-to-one from $D(B)$ onto $\H^{-1}(\R)$ (\citel{ZeidlerII} 
Theorem~32.G), hence there exists a unique $\wt\psi\in D(B) \subset \H^1(\R)$ 
such that \reff{e3.monot} holds. 

It remains to show that $\wt\psi$ is smooth, and that $\n = n+\wt\psi'$ remains
in the interval between $\eta_+$ and $\eta_-$ where no modification of $a$ was 
needed. Since $a(\eta) = A'(\eta) \geq \veps >0$, classical regularity theory 
for ordinary differential equations applied to \reff{e3.monot} shows that 
$\wt\psi$ is as smooth as the coefficients in the equation, namely 
$\wt\psi\in\CC^\infty(\R)$. Hence $\n=n+\wt\psi'$ solves
\[
\big[A(\n)\big]''+\fr\xi2 \n'= 0~, \quad \mbox{or equivalently } (a(\n)\n')' = 
-\frac{\xi}{2a(\n)}\,(a(\n)\n')~.
\]
Integrating this equation, we obtain for any $\xi_0 \in \R$
\beql{e3.6}
\n'(\xi)=\fr{a(\n(\xi_0))\n'(\xi_0)}{a(\n(\xi))}
\exp\left(-\int_{\xi_0}^\xi \fr{y}{2a(\n(y))}\, \d y\right)~, \quad 
\xi \in \R~.
\eeq
{}From this representation, it is obvious that $\n'(\xi_0)$ cannot vanish
unless $\n$ is identically constant, hence $\n$ is a monotone function
of $\xi \in \R$. In addition, since $a(\n) \geq \veps>0$, it follows from
\reff{e3.6} that $\n'$ converges to $0$ faster than any exponential 
as $\xi\to\pm\infty$. Therefore, $\n$ converge to some limits as $\xi\to \pm 
\infty$, and these limits must be the boundary value $\eta_\pm$ prescribed 
by $n$, since $\n-n = \wt\psi' \in \L^2(\R)$. Thus, parts (a) and (b) are 
established. The relations in (c) follow from \reff{e3.4} by partial 
integration, taking care of the jump of $\eta_\infty$ at $\xi=0$. Finally,
${\mathrm{sign}}(\phi_*) = {\mathrm{sign}}(\eta_+{-}\eta_-)$ since $\n$ is 
strictly monotone if $\eta_+ \neq \eta_-$.  
\QQQ

Strictly speaking, the remainder of this section is not needed for the proof
of the main results as given in Section \ref{sect.results}. However, the
statements and proofs given here for the slaved problem \reff{e3.1} help to
understand the analysis of the more difficult full problem \reff{e2.12}. 
Moreover, the arguments here show that our decay results are in a certain 
sense optimal.
   
In the above theorem we used the fact that the steady part of \reff{e3.1} has
a monotone structure. This in fact implies that the dynamics of 
\reff{e3.1} is rather trivial: all solutions are contracted 
with prescribed rate towards the unique steady state $\psi = 0$. 
Let\labell{prop3.2} $\psi_1,\psi_2$ be classical solutions of \reff{e3.1} 
in $\H^2(\R)$ satisfying $\n(\xi)+\psi_j'(\tau,\xi) \in \INTER$ for all 
$\tau\geq 0$ and $\xi\in\R$. Then
\beql{e3.2}
 \|\psi_1(\tau,\cdot)-\psi_2(\tau,\cdot)\|_{\L^2(\R)}\leq
 \e^{-3\tau/4}\|\psi_1(0,\cdot)-\psi_2(0,\cdot)\|_{\L^2(\R)}\,,
\eeq
for all $\tau \geq 0$. This follows from the energy estimate
\[\ba{rcl}
 \fr12\fr{d}{dt}\int_\R(\psi_1{-}\psi_2)^2\d\xi &=&
 \int_\R(\psi_1{-}\psi_2)\big[A(\n{+}\psi'_1)'{-}A(\n{+}\psi'_2)'\big]\,\d\xi\\
 &&+\int_\R(\psi_1{-}\psi_2)\big[\fr\xi2(\psi'_1{-}\psi'_2)-
 \fr12(\psi_1{-}\psi_2)\big]\,\d\xi \\
 &=&-\int_\R(\psi'_1{-}\psi'_2)\big[A(\n{+}\psi'_1){-}A(\n{+}\psi'_2)\big] 
 \,\d\xi -\fr34\int_\R(\psi_1{-}\psi_2)^2 \,\d \xi\,.
\ea\]
The first integral in the last row is nonnegative since $A$ is monotone,
namely, $A'(\eta)=a(\eta)>0$ implying $(\eta_1{-}\eta_2)(A(\eta_1){-}
A(\eta_2))\geq 0$. Applying Gronwall's inequality, we obtain \reff{e3.2}.

We now consider the linearization of the phase diffusion equation \reff{e3.1}
at the origin, and we discuss the spectral properties of the corresponding
differential operator $\LL:\psi\mapsto \big[a(\n)\psi'\big]'+\fr\xi2\psi'-
\fr12\psi$. According to \reff{e3.2}, the spectrum of $\LL$ in 
$\L^2(\R)$ is contained in the left half-plane $\Omega_0 = \set{z \in \C}{ 
\Re(z) \leq -3/4}$. In fact, our next result will show that $\sigma(\LL) = 
\Omega_0$ in $\L^2(\R)$, so that the estimate \reff{e3.2} cannot be improved. 
However, the upper bound of the spectrum depends on the function space, and 
can be shifted to the left if we impose a faster decay at infinity. To see 
this, we introduce for $\ell \geq 0$ the weighted space $X^\ell = 
\L^2(\R,(1+\xi^2)^\ell\d\xi)$ equipped with the scalar product $\langle \psi,
\phi \rangle = \int_\R \psi\phi (1+\xi^2)^\ell\d\xi$. Then the following 
holds:

\bprop
\labell{prop3.3}%
Let $\ell\geq 0$ and let $\LL : D(\LL) \subset X^\ell \to 
X^\ell$ be the linear operator defined by $D(\LL) = \set{\psi \in X^\ell }{
\psi'' \in X^\ell \,,\, \xi\psi' \in X^\ell }$ and $\LL\psi = (a(\n)\psi')'
+\fr\xi2\psi'-\fr12\psi$ for all $\psi \in D(\LL)$. Then there exists a 
sequence $\{\lambda_n\}_{n \ge 1} \subset \R_-$ independent of $\ell$
such that the spectrum of $\LL$ is 
\[
   \sigma(\LL) = \set{\lambda_n }{ n \ge 1 } \cup \set{z \in \C }{ 
   \Re(z) \leq -3/4-\ell/2 }\,.
\]
In addition, $\lambda_1 = -1$, $\lambda_{n{+}1} < \lambda_n$ for all 
$n \ge 1$, $\lambda_n \to -\infty$ as $n \to +\infty$, and $\lambda_n$ is a 
simple (isolated) eigenvalue of $\LL$ for all $n$ such that $\lambda_n > -3/4 
-\ell/2$. 
\eprop

\brems\\
{\bf 1)} In the case where $\eta_+ = \eta_- = \n$, it is known that $\lambda_n 
= -(n{+}1)/2$ for all $n \ge 1$, see for instance \citel{GaRau}. The
remarkable 
fact that $\lambda_1 = -1$ is the largest isolated eigenvalue of $\LL$ 
for all $\eta_+,\eta_-$ originates in the translation invariance in time 
of the Ginzburg-Landau equation \reff{e2.1}, see the discussion at the end of 
this section. \\
{\bf 2)} Proposition~\ref{prop3.3} gives the upper bound 
$\max\{-1,-3/4-\ell/2\}$ on the real part of the spectrum of $\LL$ 
in $X^\ell$. Thus, even for rapidly decreasing initial data, the solutions
of \reff{e3.1} do not converge to zero faster than $e^{-\tau}$ as $\tau \to 
+\infty$. 
\erem
\PPP
Throughout the proof, we denote by $\Omega_\ell$ the half-plane $\set{z \in \C
}{ \Re(z) \leq -3/4 -\ell/2}$. 
Let $b(\xi) = a(\n(\xi))$, and let $\gamma(\xi)$ be the solution of the 
differential equation $4b(\xi)\gamma'(\xi) + \xi\gamma(\xi) = 0$ with 
initial condition $\gamma(0) = 1$. By Theorem~\ref{thm3.1}, $b(\xi)$ 
converges (faster than any exponential) to positive limits $b_\pm$ as $\xi 
\to \pm\infty$, hence $\gamma(\xi) = \OO(\exp(-\xi^2/8b_\pm))$ as $\xi \to
\pm\infty$. 

For all $\lambda \in \C$, it is easy to verify (using asymptotic expansions)
that the linear ODE $\LL\psi = \lambda\psi$ has two independent solutions
$\psi_1^+,\psi_2^+$ satisfying
\[
  \psi_1^+(\xi) = \xi^{1{+}2\lambda}\left(1+\OO(\xi^{-2})\right)\,, \quad
  \psi_2^+(\xi) = \xi^{-2(\lambda{+}1)}\gamma(\xi)^2\left(1+\OO(\xi^{-2})
  \right)\,,
\]
as $\xi \to +\infty$. Similarly, there exist independent solutions $\psi_1^-,
\psi_2^-$ such that
\[
  \psi_1^-(\xi) = |\xi|^{1{+}2\lambda}\left(1+\OO(|\xi|^{-2})\right)\,, \quad
  \psi_2^-(\xi) = |\xi|^{-2(\lambda{+}1)}\gamma(\xi)^2\left(1+\OO(|\xi|^{-2})
  \right)\,,
\]
as $\xi \to -\infty$. Therefore, if $\Re(\lambda) < -3/4-\ell/2$, any solution
$\psi$ to $\LL\psi = \lambda\psi$ belongs to $X^\ell$, so that $\lambda$ is 
a double eigenvalue of $\LL$. This shows that $\sigma(\LL) \supset
\Omega_\ell$. 

On the other hand, if $\lambda$ is any eigenvalue of $\LL$ with $\Re(\lambda)
> -3/4-\ell/2$, then (by the same argument) $\lambda$ is simple and the 
corresponding eigenvector $\psi$ decays like $|\xi|^{-2(\lambda{+}1)}
\gamma(\xi)^2$ as $\xi \to \pm\infty$. Thus, setting $\phi(\xi) = 
\gamma(\xi)^{-1} \psi(\xi)$, we see that $\phi \in \H^2(\R)$ is a solution 
to $\Lambda\phi = \lambda\phi$, where
\[
  \Lambda\phi = (b\phi')' - V\phi\,, \quad \mbox{and } V(\xi) = 
  \fr34 + \fr{\xi^2}{16b(\xi)}\,.
\]
Since $b(\xi) \to b_\pm$ and $V(\xi) \to +\infty$ as $\xi \to \pm\infty$, 
it is well-known (see for instance \citel{CodLev}, chapter~9) that $\Lambda$ 
is self-adjoint in $\L^2(\R)$, and that its spectrum consists of a sequence 
$\{\lambda_n \}_{n \ge 1}$ of real, simple eigenvalues satisfying 
$\lambda_{n{+}1} < \lambda_n$ and $\lambda_n \to -\infty$ as $n \to +\infty$.
Moreover, the eigenfunction $\phi_n$ associated with $\lambda_n$ has exactly 
$n$ zeros in $\R$. This proves that, for all $n \ge 1$, $\lambda_n$ is an 
eigenvalue of $\LL$ with eigenfunction $\psi_n = \gamma \phi_n$, and 
that the sequence $\{\lambda_n\}_{n \ge 1}$ exhausts the discrete spectrum 
of $\LL$ in the half-plane $\Omega_\ell^c$. Now, a direct calculation shows 
that $\LL\hat\psi = -\hat\psi$, where $\hat\psi(\xi) = 2b(\xi)\n'(\xi)$ if 
$\eta_+ \neq \eta_-$ and $\hat\psi(\xi) = \exp(-\xi^2/4b)$ if $\eta_+ = 
\eta_-$. Since $\hat\psi$ has no zero in $\R$, it follows that $\lambda_1 = 
-1$ and $\psi_0 = \hat\psi$. 

It remains to verify that $\LL$ has no essential spectrum in $\Omega_\ell^c$
(in other words, if $\lambda \in \sigma(\LL) \cap \Omega_\ell^c$, then 
$\lambda$ is an isolated eigenvalue of $\LL$ with finite algebraic 
multiplicity.) To prove this, we use the change of variables defined by
$  y = B(\xi) = \int_0^\xi b(z)^{-1/2}\,\d z$, $\psi(\xi) = 
\chi(B(\xi))$. Then $(\LL\psi)(\xi) = (\LLL\chi)(B(\xi))$, where
\[
  (\LLL\chi)(y) = \chi''(y) + \fr{y}{2}\chi'(y) - \fr12\chi(y) + D(y)\chi'(y)
  \equiv (\LL_0\chi)(y) + D(y)\chi'(y)\,,
\]
and 
\[
  D(y) = \left(\fr12 b'(\xi) + \xi b(\xi)^{-1/2} - B(\xi)\right)
  \Big|_{\xi = B^{-1}(y)}\,.
\]
{}From \citel{GaRau}, we know that $\sigma_\ess(\LL_0) = \Omega_\ell$ in 
$X^\ell$. On the other hand, it follows from the Rellich criterion 
(\citel{ReedSim}, theorem XIII.65) that the operator $\partial_y \LL_0^{-1}$ 
is compact in $X^\ell$. Indeed, there exists a constant $C > 0$ such that, 
for all $\psi \in X^\ell$ with $\|\psi\|_{X^\ell} \leq 1$, the function 
$\chi = \partial_y (\LL_0^{-1}\psi)$ satisfies $\|\chi\|_{X^\ell} \leq C$, 
$\|y \chi\|_{X^\ell} \leq C$, $\|\chi'\|_{X^\ell} \leq C$. Since the function 
$D(y)$ is bounded, we conclude that $\LLL$ is a relatively compact 
perturbation of $\LL_0$. Now, it follows from a classical theorem 
(\citel{Henry}, theorem~A.1, p.~136) that either $\sigma_\ess (\LLL) = 
\sigma_\ess(\LL_0) = \Omega_\ell$, or the half-plane $\Omega_\ell^c$ is
filled with eigenvalues of $\LLL$. The second possibility is of course 
excluded, hence we have shown that $\sigma_\ess(\LL) \equiv \sigma_\ess(\LLL)
= \Omega_\ell$. 
\QQQ

Finally, we show that the eigenvalue $\lambda_1 = -1$ of $\LL$ is 
related to the symmetry properties \reff{e2.3} of the Ginzburg-Landau 
equation \reff{e2.1}. It is advantageous to discuss these symmetries for the 
slaved equation for $\eta$ given by
\beql{e3.7}
 \dot\eta= \big[A(\eta)\big]''+\fr\xi2\eta'\,.
\eeq
As was already mentioned, the phase rotation $P_\alpha$ leaves $\eta$
invariant, but the symmetries of time translation $T_z$ and space translation
$S_y$ as given in \reff{e2.7} are nontrivial. In addition, \reff{e3.7} is 
autonomous, a symmetry which was not present in the original system 
\reff{e2.12}, due to the factor $e^\tau$ in the right-hand side. 
Therefore, if $\eta(\tau,\xi)$ is a solution of \reff{e3.7}, then for 
every $\tau_0,y,z\in\R$ the function
\[
 \wh\eta_{\tau_0,y,z}(\tau,\xi)=\eta\left(\tau{+}\log(1{+}\e^{-\tau}z)+\tau_0,
 (\xi{+}\e^{-\tau/2}y)/\sqrt{1{+}\e^{-\tau}z}\right)
\]
is also a solution. 

If $\eta_+ \neq \eta_-$ we may choose for $\eta$ the steady state
$\n$ and obtain a two--dimensional invariant manifold
\[
{\mathcal M}=\set{\n((\cdot + y)/\sqrt{1{+}z\,})}{y\in \R,\; z>-1} 
\]
which is invariant under the flow of \reff{e3.7}. Using $y$ and $z$ as
coordinates on this manifold the reduced flow is $\dot y=-\fr12 y$ and
$\dot z=-z$. As a consequence, if $\KKK:\delta\mapsto
\big[a(\n(\xi))\delta]''+ \fr\xi2\delta'$ is the linearization of \reff{e3.7}
at the steady state $\n$, we have
\[
  \KKK\n'=-\fr12\n'\,, \quad\mbox{ and }\quad\KKK(\xi\n')=-(\xi\n')\,,
\]
since $\pl_y[S_y \n]\Big|_{y=0} = \e^{-\tau/2}\n'$ and $\pl_z[T_z \n]
\Big|_{z=0} = \e^{-\tau}(-\fr12\xi\n')$.

Thus, $\lambda_0=-1/2$ and $\lambda_1 = -1$ are eigenvalues of $\KKK$ for
all  $\eta_+,\eta_- \in \INTER$. Since the operators $\KKK$ and $\LL$ are
related by  $\big[\LL\psi\big]'=\KKK\psi'$, it follows that $\LL\n =
-\fr12\n$ and $\LL 
\psi_0 = - \psi_0$, where $\psi_0(\xi)=2a(\n(\xi))\n'(\xi)$ is the 
primitive of $-\xi\n'$. This explains the eigenvalue $\lambda_1 = -1$ of 
$\LL$ in the general case $\eta_+ \neq \eta_-$ (Note, however, that $-1/2$ is 
not an eigenvalue of $\LL$, since $\n$ does not decay at infinity.) 
In addition, using Proposition~\ref{prop3.3} and the relation 
$\big[\LL\psi\big]'=\KKK\psi'$, it is easy to show that the spectrum of $\KKK$
in the weighted space $X^\ell$ is given by 
$\sigma(\KKK) = \set{\lambda_n }{ n \ge 0} \cup \set{z \in \C }{ \Re(z) \leq 
-1/4-\ell/2}$. 

In fact, the global manifold $\mathcal M$ above is a special case of the
local spectral manifolds as constructed in \citel{Wayne}. This recent theory
for partial differential equations on unbounded domains uses rescaling and
weighted norms to generate discrete spectra which then allows for a
separation of the spectrum into a finite dimensional part and the remainder. 
Combining the arguments there and our Proposition \ref{prop3.3} it should be
possible to show that for all $\ell \geq 0$ and all initial data $\eta_0$ 
such that $\mathrm{dist}_{X^\ell}(\eta_0,{\mathcal M})$ is sufficiently 
small, there exist $y,z \in \R$ such that the solution $\eta$ of 
\reff{e3.7} with $\eta(0,\cdot)=\eta_0$ satisfies
\[
 \|\eta(\tau,\cdot)-\n( 
 (\cdot+\e^{-\tau/2}y)/\sqrt{1{+}\e^{-\tau}z}\,)\|_{X^\ell}=
 \OO(\e^{-\alpha\tau)}\,,
\]
as $\tau \to +\infty$, for all $\alpha<\min\{ -\lambda_2,1/4+\ell/2\}$. 
Clearly, only the case $\ell>3/4$ gives interesting results, since $\alpha>1$ 
is necessary to isolate the coordinate $z$ properly. 


\SECTION{Convergence results}
\labell{sect.results}

This section contains the convergence results for the diffusive mixing 
solutions. We first state our main result for the $(\psi,s)$ system
\reff{e2.12}, and then we transform it back to the original equation 
\reff{e2.1}. 

\bsatz
Let \labell{thm4.1}$\eta_+, \eta_- {\in}\INTER$, $\eta_+ \neq \eta_-$, and
let $\n$ be given by Theorem~\ref{thm3.1}.
There exist  $\epsilon_0 > 0$ and $\tau_0 > 0$ such that, for all 
$(\psi_0,s_0) \in \H^2(\R) \ti \H^1(\R)$ satisfying $\|\psi_0\|_{\H^2} 
+ \|s_0\|_{\H^1} \le \epsilon_0$, the system \reff{e2.12} has a unique 
global solution $(\psi,s) \in \CC([\tau_0,\infty),\H^2(\R) \ti \H^1(\R))$ 
satisfying $(\psi(\tau_0),s(\tau_0)) = (\psi_0,s_0)$. Moreover, for all
$\gamma \in (0,3/4)$ we have  
\beql{e4.dec}
\|\psi(\tau,\cdot)\|_{\H^2}=\OO(\e^{-\gamma\tau})
\quad \mbox{ and }\quad \|s(\tau,\cdot)\|_{\H^1} = \OO(\e^{-\tau})
\quad\mbox{as }\tau \to +\infty\, . 
\eeq
\esatz

We shall prove Theorem~\ref{thm4.1} in several steps. In 
Section~\ref{sect.local}  we prove a local existence result for the 
solution $(\psi,s)$ of the system \reff{e2.12} in the space $\H^2(\R) \ti 
\H^1(\R)$. In Section~\ref{sect.energy}  we introduce several energy 
functionals to control the behavior of the solution on the interval of 
existence. We derive differential inequalities for these functionals which 
imply, in particular, that for large $\tau_0$ every solution arising from 
sufficiently small data at time $\tau = \tau_0$ exists and remains bounded 
for all subsequent times. Then, using the differential inequalities again, 
we show that the solution converges exponentially to zero as 
$\tau \to +\infty$. 

These estimates are rather intricate for two  reasons. Firstly, our system is
nonautonomous through the factor $\e^\tau$ in the equation for $s$. This
factor, which amounts to an exponentially growing damping, is very essential
since it is responsible for the slaving of the amplitude $\rho$ to the phase
derivative $\eta=\n{+}\psi'$. Note that the right-hand side of the equation 
for $s$ does not vanish for $(\psi,s)=0$, hence the origin is only an 
asymptotic fixed point of \reff{e2.12}. Secondly, as was already mentioned in 
Section~\ref{sect.reform}, there is an imbalance in the number of spatial
derivatives in \reff{e2.12} which forces us to study this system in the 
rather unusual space $\H^2(\R)\ti\H^1(\R)$. In this space, \reff{e2.12} 
behave like a quasilinear system, although the original equation \reff{e2.1}
is semilinear. 

We now translate Theorem~\ref{thm4.1} back to the original variable 
$u =\sqrt{1{-}(\n{+}\psi')^2} \,\e^{\sqrt t\,[\wtN +\psi]}$. Since the 
function $u$ does not decay as $|x|\to \infty$, it is convenient to use the 
uniformly local Sobolev spaces $\Hloc{k}$, see \citel{MiSchn,MiCGL}. These 
function spaces are equipped with two different norms: a weighted norm 
($\H^k_w$ norm) and a uniformly local norm ($\H\lu^k$ norm). Without loss of 
generality, we choose here the weight function $w(x)=1/(1{+}x^2)$. For $k 
\in \N$, we define 
\beql{e4.weight}
 \|u\|_{\H^k_w}^2 = \int_\R w(x)(|u(x)|^2 {+} \dots {+} |\partial_x^k 
 u(x)|^2)\,\d x\,, \quad
 \|u\|_{\H\lu^k} = \sup_{y\in\R}\|u(\cdot{+}y)\|_{\H^k_w}\,. 
\eeq
In particular, one has $\CC^1_{\mathrm b}(\R) \subset \Hloc{1} \subset
\L^\infty(\R)$ and $\|u\|_{\L^\infty} \leq C \|u\|_{\Hloc{1}}$ for some 
$C > 0$.  

For all $u_0\in \Hloc{k}$, $k\in\N$, the Ginzburg-Landau equation 
\reff{e2.1}, has a unique global solution $u\in \CC([0,\infty),\Hloc{k})$ 
with $u(0,\cdot)=u_0$ (see \citel{Collet,MiSchn}). We now consider special 
initial conditions which are close to the mixing profiles \reff{e1.profile}
associated with $\n$. Our result is

\bsatz
Let\labell{thm4.2} $\eta_+, \eta_- {\in}\INTER$, $\eta_+ \neq \eta_-$, and 
let $\wt U$ be given by \reff{e1.profile}. There exist $t_0>0$ and $\veps>0$ 
such that, for all $u_0\in \Hloc{2}$ satisfying $\|u_0-\wt U(t_0,\cdot)
\|_{\H^2}\leq \veps$, the unique solution $u$ of \reff{e2.1} in $\Hloc{2}$ 
with $u(0,\cdot) = u_0$  satisfies, for all $\nu \in (0,1)$, 
\beql{e4.uuu}
  \|u(t,\cdot)-\wt U(t_0{+}t,\cdot)\|_{\H\lu^1} = \OO(t^{-\nu/4})\,, \quad
  \|\,|u(t,\cdot)|-|\wt U(t_0{+}t,\cdot)|\,\|_{\H\lu^1}  = 
  \OO(t^{-3\nu/4})\,,
\eeq
as $t \to +\infty$. 
\esatz

\brems\\
{\bf 1)} The loss of regularity from $\H^2$ to 
$\H^1$ is generated by the imbalance in the number of derivatives of 
$\psi$ and $s$ in \reff{e2.12}: we need an $\H^2$ condition on the data to 
ensure that $\psi \in \H^2$, but since $s \in \H^1$ we can only recover an
$\H^1$ regularity in our results. 
\\
{\bf 2)}
The initial perturbations are assumed to be small in the $\H^2$ norm, but our
convergence results are formulated in the uniformly local norm
$\|\cdot\|_{\H\lu^1}$ only, because we want to keep the optimal decay rate in
time. Indeed, since the estimates for $(\psi,s)$ are obtained in $\H^k$ norms
with respect to the diffusive variable $\xi=x/\sqrt t$
(Theorem~\ref{thm4.1}), we lose a factor $t^{1/4}$ when transforming the
results back to the corresponding $\H^k$ norms with respect to the original
variable $x$. For instance, we find  $\|\, |u(t,\cdot)|-|\wt
U(t_0{+}t,\cdot)|\, \|_{\H^1} = \OO(t^{-\nu/2})$, since the amplitudes are
just rescaled but do not involve a prefactor $\sqrt{t}$ like for 
the phase $\phi$. However, it is not possible to give a definite decay rate
for $\|u(t,\cdot)-\wt U(t_0{+}t,\cdot)\|_{\H^1}$ 
unless we impose further spatial decay properties. Assuming the perturbation
to lie in $X^\ell$ (together with the first two derivatives) it is possible to
improve the decay in \reff{e4.dec} to $\OO(\e^{-\alpha \tau})$ for all $\alpha
<\min\{1,3/4{+}\ell/2\}$ (see \citel{BricKup} and the arguments at the end 
of Section~\ref{sect.phase} above). Then $\|u(t,\cdot){-}\wt U(t_0{+}t,\cdot)
\|_{\H^1} = \OO(t^{3/4-\alpha})$ can be concluded, which gives a positive 
decay rate whenever $\ell>1/2$. 
\erem
\PPP
Let $t_0 = e^{\tau_0}$, where $\tau_0$ is given by Theorem~\ref{thm4.1}. 
If $u_0 \in \Hloc{2}$ satisfies $\|u_0 - \wt U(t_0,\cdot)\|_{\H\lu^2} \leq
\veps$ for some sufficiently small $\veps > 0$, there exists a unique pair
of functions $(\psi_0,s_0) \in \H^2(\R)\ti\H^1(\R)$ such that
\[
 u_0(\sqrt{t_0}\xi) = \sqrt{1{-}(\n(\xi){+}\psi_0'(\xi))^2}\,\e^{s_0(\xi)/2}
 \,\e^{\i \sqrt{t_0}\,[\wtN(\xi)+\psi_0(\xi)]}\,, \quad \xi \in \R\,.
\]
In addition, one has $\|\psi_0\|_{\H^2} + \|s_0\|_{\H^1} \leq C \veps$ 
for some $C > 0$. Thus, if $C\veps \leq \veps_0$, Theorem~\ref{thm4.1}
shows that the unique solution of \reff{e2.12} in $\H^2(\R)\ti\H^1(\R)$ 
with initial data $(\psi_0,s_0)$ satisfies $\|\psi(\tau,\cdot)\|_{\H^2} + 
\|s(\tau,\cdot)\|_{\H^1} = \OO(\e^{-\gamma \tau})$ as $\tau \to +\infty$, 
for all $\gamma < 3/4$. 
In particular, we have for all $\nu < 1$:
\[\ba{l}
 \int_\R w(x)(\psi^2{+}\psi'^2{+}s^2)\bigl(x/\sqrt t,\log t\bigr)\d x
  \leq C \sup_{\xi \in \R} (\psi^2{+}\psi'^2{+}s^2)(\xi,\log t) = 
  \OO(t^{-3\nu/2})\,,\\
 \int_\R w(x)(\psi''^2{+}s'^2)\bigl(x/\sqrt t,\log t\bigr)\d x \leq
  C \sqrt{t}\int_\R (\psi''^2+s'^2)(\xi,\log t)\d\xi = \OO(t^{-\nu})\,,
\ea
\]
where $w$ is the weight function $1/(1{+}x^2)$ or any translate of it. 
Using these results, it is straightforward to verify that the unique 
solution $u(t,x)$ of \reff{e2.1} in $\Hloc{2}$ with initial data $u_0$
is given by
\[
u(t,x)=\sqrt{1{-}(\n(\xi){+}\psi'(\tau,\xi))^2}\,\e^{s(\tau,\xi)/2}
\, \e^{\i \sqrt t\,[\wtN(\xi) +\psi(\tau,\xi)]}\,,\quad
\mbox{with } \tau=\log (t{+}t_0)\,,~\xi=\fr{x}{\sqrt{t{+}t_0}}\,,
\]
and that the decay estimates \reff{e4.uuu} are satisfied. 
\QQQ

Theorem~\ref{thm4.2} implies that, up to a time-dependent phase, the 
solution $u(t,x)$ converges uniformly on compact sets to the stationary
solution $\sqrt{1-\eta_*^2}\,\e^{\i \eta_* x}$ as $t \to +\infty$, where
$\eta_* = \n(0)$. Indeed, setting $\phi_* = \wtN(0) = \int_{-\infty}^0 
(\n(\xi)-\eta_-)\d\xi \neq 0$ as in Theorem~\ref{thm3.1}, we have the 
following result (see also \citel{BricKup}):

\bcoro
Under\labell{cor4.3} the assumptions of Theorem~\ref{thm4.2}, we have for 
all $x_0>0$ and all $\nu \in (0,1)$ the estimate
\beql{e4.bric}
 \sup_{|x|\leq x_0}|u(t,x)-\sqrt{1{-}\eta_*^2}\,
 \e^{\i[\sqrt t\,\phi_*+\eta_*x]}|
 \,=\, \OO(t^{-\nu/4})\,, \quad t \to +\infty\,,
\eeq
where $\eta_* = \n(0)$ and $\phi_* = \wtN(0)$. 
\ecoro
\PPP 
Since $\n(x/\sqrt t)=\eta_*+\OO(t^{-1/2})$ and $\sqrt t\, \wtN(x/\sqrt t)
= \sqrt t \phi_* + \eta_* x + \OO(t^{-1/2})$ uniformly for $|x| \leq x_0$, 
this result follows immediately from the first estimate in \reff{e4.uuu}. 
\QQQ

\brem 
In the case where both $\eta_+, \eta_-$ are close to zero, the estimate 
\reff{e4.bric} is obtained in \citel{BricKup} with the better decay rate
$\OO(t^{-\nu/2})$ for the remainder. As was already mentioned in the 
previous remark, this is because we allow for initial perturbations with 
only $\L^2$ decay at infinity, while in \citel{BricKup} stronger norms are 
used which imply in particular that the perturbations lie in $X^\ell$ 
with $\ell\geq 1$. 
\erem


\SECTION{Local existence}
\labell{sect.local}

In this section, we prove a local existence result for the solutions 
$(\psi,s)$ of the system \reff{e2.12} in the space $\H^2(\R)\ti\H^1(\R)$.
Although it originates from the simple Ginzburg-Landau equation \reff{e2.1}, 
this system in nonautonomous due to the change of variables \reff{e2.5}. 
In addition, the nonlinear transformation \reff{e2.9} forces us to study
it in the unbalanced space $\H^2(\R)\ti\H^1(\R)$, where it behaves like a 
quasilinear system. For these reasons, we have to take special care of the 
regularity of the solutions and the length of the local existence intervals.

Throughout this section, we assume that $\eta_+,\eta_- \in \INTER$ and that
$\n$ is given by Theorem~\ref{thm3.1}. Our result is:

\bprop
There\labell{prop5.1} 
exist $\epsilon_1 > 0$, $T_1 > 0$, and $K_1 \ge 1$ such that, for all 
$\tau_1 \ge 0$ and all $(\psi_1,s_1) \in \H^2(\R) \ti \H^1(\R)$ such that
$\|\psi_1\|_{\H^2} + \|s_1\|_{\H^1} \le \epsilon_1$, the system 
\reff{e2.12} has a unique solution $(\psi,s) \in \CC([\tau_1,\tau_2],\H^2(\R)
\ti \H^1(\R))$ satisfying $(\psi(\tau_1),s(\tau_1)) = (\psi_1,s_1)$,  
where $\tau_2 = \tau_1 + \log(1+T_1 \e^{-\tau_1})$. This solution depends
continuously on the initial data $(\psi_1,s_1)$ in $\H^2(\R) \ti \H^1(\R)$, 
uniformly in $\tau \in [\tau_1,\tau_2]$. Moreover, the bound
\beql{e4.1}
  \|\psi(\tau)\|_{\H^2} + \|s(\tau)\|_{\H^1} \,\le\, 
  K_1 \left(\|\psi_1\|_{\H^2} + \|s_1\|_{\H^1} + \e^{\tau_1}(\tau-\tau_1)
  \right)
\eeq
holds for all $\tau \in [\tau_1,\tau_2]$. 
\eprop

\brems\\
{\bf 1)} Here and in the sequel, we simply write $\psi(\tau), s(\tau)$ 
instead of $\psi(\tau,\cdot), s(\tau,\cdot)$ when no confusion is possible.\\
{\bf 2)} In particular, Proposition~\ref{prop5.1} implies that, if 
$(\psi,s) \in \CC([\tau_1,\tau^*),\H^2(\R)\ti\H^1(\R))$ is a maximal solution
of \reff{e2.12} which satisfies $\|\psi(\tau)\|_{\H^2} + \|s(\tau)\|_{\H^1}
\leq \epsilon_1$ for all $\tau \in [\tau_1,\tau^*)$, then actually $\tau^* = 
+\infty$, {\sl i.e.} the solution can be continued to the whole interval
$[\tau_1,+\infty)$.\\
{\bf 3)} The proof shows that the solution $\psi(\tau,\xi), s(\tau,\xi)$
of \reff{e2.12} is a $\CC^\infty$ function of $\tau, \xi$ for all $\tau > 
\tau_1$. However, it is not true in general that $(\psi,s) \in 
\CC^k((\tau_1,\tau_2],\H^2(\R) \ti \H^1(\R))$ for $k > 0$, unless we assume 
in addition that $\psi_1(\xi), s_1(\xi)$ decay sufficiently fast as $|\xi| 
\to \infty$. 
\erem
\PPP 
Instead of working directly on \reff{e2.12}, we shall use the change 
of variables \reff{e2.5a}, \reff{e2.9} and solve the corresponding initial
value problem for the simpler system \reff{e2.4}. Let $\tau_1 \ge 0$, and 
let $(\psi_1,s_1) \in \H^2(\R) \ti \H^1(\R)$ be initial data for \reff{e2.12}
at time $\tau = \tau_1$ satisfying $\|\psi_1\|_{\H^2} + \|s_1\|_{\H^1} = 
\epsilon$ for some $\epsilon \le 1/4$. The corresponding initial data for 
\reff{e2.4} are given by
\beql{e4.2}
  \phi(t_1,x) \,=\, \phi_L(x) + \bar \phi_1(x)~, \quad
  r(t_1,x) \,=\, r_L(x) + \bar r_1(x)~, 
\eeq
where $t_1 = \e^{\tau_1}$, $L = \e^{\tau_1/2}$, $\phi_L(x) = L\wtN(x/L)$, 
$r_L(x) = (1-\n(x/L)^2)^{1/2}$, and
\[
  \bar \phi_1(x) \,=\, L \psi_1(x/L)~, \quad
  \bar r_1(x) \,=\, \left(1-(\n(x/L){+}\psi_1'(x/L))^2\right)^{1/2}
  \e^{\frac{1}{2}s_1(x/L)} - r_L(x)~.
\]
A direct calculation shows that there exists a constant $C_1 > 0$ (independent
of $L$) such that $\|\bar \phi_1\|_{\H^2_L} + \|\bar r_1\|_{\H^1_L} \le C_1 
(\|\psi_1\|_{\H^2} + \|s_1\|_{\H^1})$, where $\H^2_L$, $\H^1_L$ are the 
Sobolev spaces $\H^2(\R)$, $\H^1(\R)$ equipped with the $L-$dependent norms
\beql{e4.3}
  \|\bar \phi\|_{\H^2_L}^2 \,=\, L^{-3} \|\bar \phi\|_{\L^2}^2 + 
    L^{-1} \|\bar \phi'\|_{\L^2}^2 + L \|\bar \phi''\|_{\L^2}^2~, \quad 
  \|\bar r\|_{\H^1_L}^2 \,=\, L^{-1} \|\bar r\|_{\L^2}^2 + 
    L \|\bar r'\|_{\L^2}^2~.
\eeq

Motivated by \reff{e4.2}, we look for a solution of \reff{e2.4} of the form
$\phi(t,x) = \phi_L(x) + \bar \phi(t,x)$, $r(t,x) = r_L(x) + \bar r(t,x)$, 
where $\bar \phi, \bar r$ satisfy the evolution system
\beql{e4.4}\ba{rcl}
  \partial_t \bar \phi &=& \partial_x^2 \bar \phi + \phi_L'' + 
   2 (r_L{+}\bar r)^{-1}(r_L'{+}\partial_x \bar r)(\phi_L'{+}\partial_x 
   \bar \phi)~, \\
  \partial_t \bar r &=& \partial_x^2 \bar r + r_L'' - (\bar r(r_L{+}2\bar r)
   + (\partial_x \bar \phi)(2\phi_L'{+}\partial_x \bar \phi))(r_L{+}\bar r)~,
\ea\eeq
together with the initial condition $(\bar \phi(t_1),\bar r(t_1)) = (\bar 
\phi_1, \bar r_1) \in \H^2_L\ti\H^1_L$. If $C_1 \epsilon \le 1/4$,
Lemma~\ref{lem5.2} below shows that this initial value problem
has a unique solution $(\bar \phi,\bar r) \in \CC([t_1,t_1{+}T],\H^2_L \ti 
\H^1_L)$, for some $T > 0$ (independent of $L$). This solution depends 
continuously on the initial data $(\bar \phi(t_1),\bar r(t_1))$ in $\H^2_L \ti 
\H^1_L$, uniformly in $t \in [t_1,t_1{+}T]$. In addition, there exist 
$C_2 > 0$, $C_3 \ge 1$ (independent of $L$) such that
\beql{e4.4a} 
  \|\bar \phi(t)\|_{\H^2_L} + \|\bar r(t)\|_{\H^1_L} \,\le\, 
  C_2(t-t_1) + C_3 C_1 \epsilon~,
\eeq
for all $t \in [t_1,t_1{+}T]$. Finally, due to the parabolic regularization, 
$\bar \phi(t,x)$ and $\bar r(t,x)$ are $\CC^\infty$ functions of $t,x$ for 
all $t > t_1$. 

Having constructed the solution $(\bar \phi, \bar r)$ of \reff{e4.4},
we now return to the variables $(\psi,s)$ defined by the relations 
\reff{e2.5}, \reff{e2.9}. Setting $\sigma = \tau-\tau_1$ and using the 
definitions above of $L$, $\phi_L$, $r_L$, we arrive at the expressions
\beql{e4.5}\ba{rcl}
  \psi(\tau_1{+}\sigma,\xi) &=& \e^{-\sigma/2} \wtN(\xi \e^{\sigma/2}) - 
   \wtN(\xi)+ \e^{-\sigma/2} L^{-1} \bar \phi(\xi L \e^{\sigma/2}, L^2 
   \e^\sigma)~, \\
  s(\tau_1{+}\sigma,\xi) &=& 2 \log[R(\xi)^{-1}R(\xi \e^{\sigma/2})] + 
   2 \log\bigl[1 + R(\xi \e^{\sigma/2})^{-1} \bar r(\xi L \e^{\sigma/2},L^2 
   \e^\sigma)\bigr] \\
  && - \log \bigl[1 - 2R(\xi)^{-2} (\n'(\xi)\psi'(\xi,\tau){+}
   \psi'(\xi,\tau)^2)\bigr]~,
\ea\eeq
where $0 \le \sigma \le \sigma_1 = \log(1+T\e^{-\tau_1})$ and $R(\xi) = 
\sqrt{1-\n(\xi)^2}$. If we assume that $C_2 T + C_1 C_3 \epsilon \le 1/4$, 
then a direct calculation shows that $(\psi,s) \in \CC([\tau_1,\tau_1{+}
\sigma_1],\H^2 \ti \H^1)$ satisfies, for all $\tau \in [\tau_1,\tau_1{+}
\sigma_1]$,
\beql{e4.5a}
  \|\psi(\tau)\|_{\H^2} + \|s(\tau)\|_{\H^1} \,\le\, 
  C_4(\tau-\tau_1) + C_5 (\|\bar \phi(\e^\tau)\|_{\H^2_L} + 
  \|\bar r(\e^\tau) \|_{\H^1_L})~, 
\eeq
where $C_4 > 0$ and $C_5 \ge 1$ are independent of $L$ (or, equivalently, of
$\tau_1$). The proof of \reff{e4.5a} relies on the properties of $\n(\xi)$ 
listed in Theorem~\ref{thm3.1} and on the fact that the dilation $\lambda 
\mapsto \phi(\lambda\cdot)$ is a continuous operation in $\H^k(\R)$, $k \in
\N$. The uniformity in $L$ of the constants $C_4, C_5$ follows from the 
definition of the scaled norms \reff{e4.3} and from the fact that $0 \le 
\sigma \le T$ uniformly in $L$. 

By construction, $(\psi,s) \in \CC([\tau_1,\tau_1{+}\sigma_1],\H^2 \ti \H^1)$ 
is a solution of the system \reff{e2.12} satisfying $(\psi(\tau_1),s(\tau_1))
= (\psi_1,s_1)$. The uniqueness of this solution and its continuous dependence
on the data follow directly from the corresponding properties of the 
system \reff{e4.4}, see Lemma~\ref{lem5.2}. Finally, combining 
\reff{e4.4a}, \reff{e4.5a}, we obtain
\[
  \|\psi(\tau)\|_{\H^2} + \|s(\tau)\|_{\H^1} \,\le\, C_1 C_3 C_5 \epsilon 
  + (C_4 + C_2 C_5 \e^\tau)(\tau-\tau_1)~,
\]
for all $\tau \in [\tau_1,\tau_1{+}\sigma_1]$. Therefore, we see that
\reff{e4.1} holds if we set, for example, 
\[
  T_1 \,=\, \min\left(T~,~\frac{1}{8C_2}\right)~, \quad
  \epsilon_1 \,=\, \frac{1}{8C_1 C_3}~, \quad
  K_1 \,=\, \max\left(C_1 C_3 C_5~,~C_4 + C_2 C_5 (1{+}T_1)\right)~.
\] 
This concludes the proof of Proposition~\ref{prop5.1}. 
\QQQ

Proposition~\ref{prop5.1} relies on a local existence result for the 
solutions of \reff{e4.4} which is the content of the next Lemma. Since the 
system \reff{e4.4} is autonomous, we assume here (without loss of generality)
that the initial time is $t_1 = 0$. We recall that $\H^2_L$, $\H^1_L$ are the
function spaces defined by the norms \reff{e4.3}. 

\blem There\labell{lem5.2} exist $T > 0$, $C_2 > 0$, $C_3 \ge 1$ such that, 
for all $L \ge 1$ and for all $(\bar \phi_1, \bar r_1) \in \H^2_L \ti \H^1_L$
such that $\|\bar \phi_1\|_{\H^2_L} + \|\bar r_1\|_{\H^1_L} \le 1/4$, the 
system \reff{e4.4} has a unique solution $(\bar \phi, \bar r) \in 
\CC([0,T],\H^2_L \ti \H^1_L)$ satisfying $(\bar \phi(0),\bar r(0)) = 
(\bar \phi_1,\bar r_1)$. This solution depends continuously on the 
initial data $(\bar \phi_1,\bar r_1)$ in $\H^2_L \ti \H^1_L$, uniformly 
in $t \in [0,T]$. Moreover, the bound
\beql{ea.1}
  \|\bar \phi(t)\|_{\H^2_L} + \|\bar r(t)\|_{\H^1_L} 
  \,\le\, C_2 t + C_3 (\|\bar \phi_1\|_{\H^2_L} + \|\bar r_1\|_{\H^1_L})~,
\eeq
holds for all $t \in [0,T]$. 
\elem
\PPP 
Lemma~\ref{lem5.2} is a standard local existence result, except
that we have to control the dependence on the scaling parameter $L \ge 1$, 
and that we need one more derivative for $\bar \phi$ than for $\bar r$. 
Because of this imbalance, the apparently semilinear equation \reff{e4.4}
behaves in fact like a quasilinear system. In particular, the constant $C_3$
in \reff{ea.1} cannot be replaced by $1$. 

Throughout the proof, we write $(\phi,r)$ instead of $(\bar \phi,\bar r)$
for simplicity. The system \reff{e4.4} becomes
\beql{ea.2}
  \partial_t \phi \,=\, \partial_x^2 \phi + F_L(\phi,r)~, \quad
  \partial_t r \,=\, \partial_x^2 r + G_L(\phi,r)~, 
\eeq
where
\[\ba{rcl}
  F_L(\phi,r) &=& \phi_L'' + 2 (r_L + r)^{-1}(r_L'+\partial_x r)
   (\phi_L'+\partial_x \phi)~, \\
  G_L(\phi,r) &=& r_L'' - \bigl(r(r_L+2r)+(\partial_x \phi)(2\phi_L'+
   \partial_x \phi)\bigr)(r_L+r)~.
\ea\]
To avoid the difficulty related to the imbalance of derivatives, we 
first prove the existence of a unique local solution of \reff{ea.2}
in a subspace of $\H^2_L \ti \H^1_L$ in which the derivatives are balanced. 
This will be done using a standard contraction mapping argument.

Let $L \ge 1$, and let $(\bar \phi_1,\bar r_1) \in \H^2_L \ti \H^1_L$ 
satisfying $m_1 \equiv \|\bar \phi_1\|_{\H^2_L} + \|\bar r_1\|_{\H^1_L} \le 
1/4$. For any $T > 0$, we denote by $X(T)$ the Banach space
\[
  X(T) = \left\{ (\phi,r) \,\Big|\, \phi,r \in \CC([0,T],\H^1_L)~,~
  \partial_x \phi \in \CC([0,T],\L^\infty)~,~
  \partial_x^2 \phi \in \L^2([0,T],\L^2)\right\}~,
\]
equipped with the norm $\|(\phi,r)\|_{X(T)}$ given by
\[
  \max\Bigl\{ \sup_{0 \le t \le T} 
  (\|r(t)\|_{\H^1_L}+L^{-1}\|\phi(t)\|_{\H^1_L})\,,~ \sup_{0 \le t \le T} 
  \|\partial_x\phi(t)\|_{\L^\infty}\,,~ \left(L\int_0^T 
  \|\partial_x^2\phi(s)\|_{\L^2}^2\,\d s\right)^{1/2}\Bigr\}\,.
\]
We note $\tilde \phi(t) = \e^{t\partial_x^2}\bar \phi_1$, $\tilde r(t) = 
\e^{t\partial_x^2}\bar r_1$, where $\e^{t\partial_x^2}$ is the heat kernel. 
Then $(\tilde \phi, \tilde r) \in X(T)$ and $\|(\tilde \phi,\tilde r)\|_{X(T)}
\le 1/4$ for any $T \le 1$. Let $\B(T)$ be the ball of radius $1/4$ centered 
at $(\tilde \phi,\tilde r)$ in $X(T)$. For all $(\phi,r) \in \B(T)$, we define
\beql{ea.hat}
  \hat \phi(t) \,=\, \int_0^t \e^{(t-s)\partial_x^2} F_L(\phi(s),
    r(s))\,\d s~, \quad
  \hat r(t) \,=\, \int_0^t \e^{(t-s)\partial_x^2} G_L(\phi(s),
    r(s))\,\d s~. 
\eeq
We shall show that $(\hat \phi,\hat r) \in X(T)$ and 
$\|(\hat \phi,\hat r)\|_{X(T)} \le 1/4$ if $T$ is sufficiently small 
(uniformly in $L$). Indeed, since $(\phi,r) \in \B(T)$, we have 
$r_L(x) + r(x,t) \ge \inf_x r_L(x) - \|r(t)\|_{\L^\infty} \ge \sqrt{2/3}-1/2 
\ge 1/4$ for all $x \in \R$, $t \in [0,T]$. As a consequence, we have
\beql{ea.3}\ba{rcl}
  \|F_L(\phi,r)\|_{\L^2} \!&\le&\! \|\phi_L''\|_{\L^2} + 8\|r_L'{+}
   \partial_x r\|_{\L^2} \|\phi_L'{+}\partial_x \phi\|_{\L^\infty} 
   \le CL^{-1/2}~, \\
  \|G_L(\phi,r)\|_{\L^2} \!&\le&\! \|r_L''\|_{\L^2} + \bigl(\|r\|_{\L^2}
   \|r_L{+}2r\|_{\L^\infty} +\|\partial_x \phi\|_{\L^2} \|2\phi_L'{+} 
   \partial_x \phi\|_{\L^\infty}\bigr) \|r_L{+}r\|_{\L^\infty} \\
  &\le& CL^{-3/2} + CL^{1/2} \le 2C L^{1/2}~,
\ea\eeq
for some $C > 0$ (independent of $L$). Therefore, using standard estimates
for the heat kernel, we obtain for all $t \in [0,T]$:
\beql{ea.4}\ba{l}
  \|\hat \phi(t)\|_{\L^2} \le C L^{-1/2} t~, \quad
   \|\partial_x \hat \phi(t)\|_{\L^2} \le CL^{-1/2} \sqrt{t}~,\quad
   \|\partial_x \hat \phi(t)\|_{\L^\infty} \le CL^{-1/2} t^{1/4}~,\\
  \|\hat r(t)\|_{\L^2} \le C L^{1/2} t~, \quad 
   \int_0^t \|\partial_x^2 \hat \phi(s)\|_{\L^2}^2\,\d s \le C L^{-1} t~.
\ea\eeq 

Moreover, differentiating $G_L$ with respect to $x$ and proceeding as above, 
we find $\|\partial_x G_L(\phi,r)\|_{\L^2} \le C L^{-1/2} + C\|\partial_x^2 
\phi\|_{\L^2}$, hence
\beql{ea.6}
  \|\partial_x \hat r(t)\|_{\L^2} \,\le\, \left\{t\int_0^t \|\partial_x
  G_L(\phi,r)\|_{\L^2}^2 \,\d s\right\}^{1/2} \!\,\le\, C L^{-1/2}t + 
  C L^{-1/2} \sqrt{t} \,\le\, C L^{-1/2} \sqrt{t}\,,
\eeq
for all $t \in [0,T]$. From \reff{ea.4}, \reff{ea.6}, we conclude that
$\|(\hat \phi,\hat r)\|_{X(T)} \le C T^{1/4}$, uniformly in $L \ge 1$. 
This proves that the mapping $M$ defined by $M(\phi,r) = (\tilde \phi + 
\hat \phi, \tilde r + \hat r)$ maps the ball $\B(T)$ into itself, 
if $T$ is sufficiently small. Using similar estimates, one shows that $M$
is a contraction in $\B(T)$, hence has a unique fixed point $(\phi,r) \in 
\B(T)$. By construction, this fixed point is the unique solution of 
\reff{ea.2} in $X(T)$ satisfying $(\phi(0),r(0)) = (\bar \phi_1,\bar r_1)$, 
and this solution depends continuously in $X(T)$ on the initial data
$(\bar \phi_1,\bar r_1) \in \H^2_L \ti \H^1_L$. 

It remains to verify that $\partial_x^2 \phi \in \CC([0,T],\L^2)$ and that
\reff{ea.1} holds. Since $(\phi,r) = (\tilde \phi, \tilde r) + (\hat \phi,
\hat r)$, where $(\hat \phi,\hat r)$ is given by \reff{ea.hat}, and since
$\|\tilde \phi(t)\|_{\H^2_L} + \|\tilde r(t)\|_{\H^1_L} \le m_1 = \|\bar 
\phi_1\|_{\H^2_L} + \|\bar r_1\|_{\H^1_L}$, it is sufficient to show that 
$\partial_x^2 \hat \phi \in \CC([0,T],\L^2)$ and $\|\hat \phi(t)\|_{
\H^2_L} + \|\hat r(t)\|_{\H^1_L} \le C(m_1+t)$ for $t \in [0,T]$. 
Appropriate bounds on $\|\hat \phi\|_{\L^2}$ and $\|\hat r\|_{\L^2}$ are 
already contained in \reff{ea.4}. To bound $\|\partial_x \hat r\|_{\L^2}$, 
we observe that 
\beql{ea.7}
  \int_0^t \|\partial_x^2 \phi(s)\|_{\L^2}^2\,\d s \,\le\, 2 \int_0^t 
  \left(\|\partial_x^2 \tilde \phi(s)\|_{\L^2}^2 + \|\partial_x^2 \hat \phi(s)
  \|_{\L^2}^2\right)\d s \,\le\, C L^{-1} t~,
\eeq
by \reff{ea.4}, hence the estimate \reff{ea.6} can be improved as follows:
\beql{ea.8}
  \|\partial_x \hat r(t)\|_{\L^2} \,\le\, \left\{t\int_0^t C(L^{-1} + 
  \|\partial_x^2 \phi(s)\|_{\L^2}^2)\,\d s\right\}^{1/2} \,\le\, C L^{-1/2} t~.
\eeq
To bound $\|\partial_x \hat \phi\|_{\L^2}$ and $\|\partial_x^2 \hat 
\phi\|_{\L^2}$, we need to estimate $\partial_x F_L$. It is convenient
here to write $F_L(\phi,r) = F_L(0,0) + \wt F_L(\phi,r)$, where $F_L(0,0) = 
\phi_L'' + 2r_L' \phi_L'/r_L$. Using \reff{ea.4}, \reff{ea.8}, it is 
straightforward to verify that
\beql{ea.5}\ba{l}
  \|\partial_x F_L(0,0)\|_{\L^2} \le C L^{-3/2}\,, \quad 
     \|\partial_x^2 F_L(0,0)\|_{\L^2} \le C L^{-5/2}\,,\\
  \|\partial_x \wt F_L(\phi,r)\|_{\L^2} \le C \|\partial_x^2 r\| + 
    C(L^{-1} + \|\partial_x r\|_{\L^\infty})(m_1 + L^{-1/2}\sqrt{t} + 
    \|\partial_x^2 \hat \phi\|_{\L^2})~.
\ea\eeq
On the other hand, using \reff{ea.4}, \reff{ea.8} as well as various standard
estimates for the heat kernel, it is not difficult to show that
\beql{ea.9}\ba{l}
  \int_0^t \|\partial_x^2 r(s)\|_{\L^2}^2 \,\d s \,\le\, CL^{-1}(m_1^2+t^2)~, 
  \quad  \int_0^t \|\partial_x r(s)\|_{\L^\infty}^2 \,\d s \,\le\, CL^{-1}
  \sqrt{t}(m_1^2+t^2)~, \\
  \int_0^t \|\partial_x r(s)\|_{\L^\infty}^2 \|\partial_x^2 \hat 
  \phi(s)\|_{\L^2}^2 \,\d s \,\le\, CL^{-2}\sqrt{t}(m_1^2+t^2)~. 
\ea\eeq
Note that the first estimate in \reff{ea.9} does not converge to zero as
$t \to 0$, since $r(0) = \bar r_1$ belongs to $\H^1_L$ only. Combining 
\reff{ea.5}, \reff{ea.9}, we thus find
\beql{ea.10}
  \int_0^t \|\partial_x \wt F_L(\phi,r)\|_{\L^2}^2 \,\d s \,\le\, 
  CL^{-1}(m_1^2+t^2) + CL^{-2} \int_0^t \|\partial_x^2 \hat \phi\|_{\L^2}^2 
  \,\d s\,.
\eeq
This result implies that $\partial_x F_L \in \L^2([0,T],\L^2(\R))$ and 
$\int_0^t \|\partial_x F_L(\phi,r)\|_{\L^2}^2 \,\d s \leq C L^{-1}
(m_1^2 + t^2) + C L^{-3}t$ by \reff{ea.4}, \reff{ea.5}. It follows that 
$\hat \phi \in \CC([0,T],\H^2_L)$ and 
\[\ba{rcl}
  \|\partial_x \hat \phi(t)\|_{\L^2} &\le& \left\{t\int_0^t \|\partial_x
    F_L(\phi,r)\|_{\L^2}^2 \,\d s\right\}^{1/2} \,\le\,
    CL^{-1/2}\sqrt{t}(m_1+t) + CL^{-3/2}t~, \\
  \|\partial_x^2 \hat \phi(t)\|_{\L^2} &\le& \left\{\int_0^t \|\partial_x
    F_L(\phi,r)\|_{\L^2}^2 \,\d s\right\}^{1/2} \,\le\, CL^{-1/2}(m_1+t)
    + CL^{-3/2}\sqrt{t}~. 
\ea\]
In particular, $\|\partial_x \hat\phi(t)\|_{\L^2} \leq CL^{-1/2}(m_1+t)$, 
which is the desired estimate. The bound on $\|\partial_x^2 \hat 
\phi\|_{\L^2}$ is not sufficient yet, but replacing it into \reff{ea.10}, we 
find $\int_0^t \|\partial_x \wt F_L(\phi,r)\|_{\L^2}^2 \,\d s \leq C L^{-1}
(m_1^2 + t^2)$, which in turn implies
\[
  \|\partial_x^2 \hat \phi(t)\|_{\L^2} \leq \int_0^t \|\partial_x^2 
  F_L(0,0)\|_{\L^2} \,\d s + \left\{\int_0^t \|\partial_x \wt 
  F_L(\phi,r)\|_{\L^2}^2 \,\d s\right\}^{1/2} \leq C L^{-1/2}(m_1+t)\,.
\]
Thus, we have shown that $\|\hat \phi\|_{\H^2_L} + \|\hat r\|_{\H^1_L} \leq 
C(m_1+t)$ for all $t \in [0,T]$, and the continuous dependence on the data 
follows by the same estimates. The proof of Lemma~\ref{lem5.2} is complete.  
\QQQ


\SECTION{Energy estimates}
\labell{sect.energy}

This final section is devoted to the proof of Theorem~\ref{thm4.1} using
energy estimates. Let $\eta_+, \eta_- \in \INTER$, $\eta_+ \neq \eta_-$, and 
assume that $(\psi,s) \in \CC([\tau_1,\tau_2], \H^2(\R) \ti \H^1(\R))$ 
is a solution of \reff{e2.12} defined on some time interval
$[\tau_1,\tau_2] \subset \R_+$ and satisfying $\|\psi(\tau)\|_{\H^2} 
+ \|s(\tau)\|_{\H^1} \le \epsilon$ for all $\tau \in [\tau_1,\tau_2]$,  
for some sufficiently small $\epsilon > 0$ (to be specified later). We 
set $(\psi_1,s_1) = (\psi(\tau_1),s(\tau_1))$. To control the evolution of 
the solutions $(\psi,s)$, we introduce the energy functionals
\beql{e4.8}\ba{rcl}
  E_0(\tau) &=& \frac{1}{2} \int_\R \psi(\tau,\xi)^2 \,\d\xi~, \\
  E_{k{+}1}(\tau) &=& \frac{1}{2} \int_\R (3\delta^{(k)}(\tau,\xi)^2 + 
   s^{(k)}(\tau,\xi)^2) \,\d\xi~,\quad k = 0,\,1,\,2\,,
\ea\eeq
where $\delta = \psi'$. 

The use of the quadratic forms $3(\delta^{(k)})^2 + (s^{(k)})^2$ instead of 
$(\delta^{(k)})^2 + (s^{(k)})^2$ in \reff{e4.8} can be understood as follows. 
For any $\beta > 0$, the system \reff{e2.12},
\reff{e2.13} for $\delta,s$ can be written in the form
\beql{e4.9}
  \pmatrix{\beta \dot \delta \cr \dot s} \,=\, 
  \pmatrix{a(\eta) & \beta \eta \cr \beta^{-1} a_2(\eta) & a_1(\eta)}
  \pmatrix{\beta \delta'' \cr s''}
  \,+\, R_\beta(\xi,\tau;\delta,\delta',s,s')~, 
\eeq
where $a(\eta), a_1(\eta), a_2(\eta)$ are defined in \reff{e2.aaa} and the 
remainder $R_\beta$ contains only lower order derivatives of $\delta,s$. 
If $M_\beta(\eta)$ denotes the $2 \ti 2$ matrix in the right-hand side of 
\reff{e4.9}, then $\Tr M_\beta(\eta) = 2$ and $\Det M_\beta(\eta) = 1$ 
for all $\beta > 0$, $|\eta| < 1/\sqrt{3}$, so that $1$ is always a double 
eigenvalue of $M_\beta$. However, the evolution of the functionals $\int
(\beta^2 (\delta^{(k)})^2 + (s^{(k)})^2)\d\xi$ is determined (up to lower 
order terms) by the symmetrized matrix $M_\beta^s = {1 \over 2}(M_\beta + 
M_\beta^\tp)$, which is not necessarily positive definite. Our choice of 
$E_{k{+}1}(\tau)$ in \reff{e4.8} is motivated by the fact that $M_\beta^s$ 
is positive definite for all $|\eta| < 1/\sqrt{3}$ if and only if $\beta = 
\sqrt{3}$. In this case, the eigenvalues are given by
\beql{e4.10}
   \mu_\pm(\eta) \,=\, 1 \pm {\sqrt{3}|\eta| \over 6(1-\eta^2)^2}
   (3-2\eta^2+3\eta^4)~.
\eeq

In the sequel, we set $\eta_0 = \max(|\eta_+|,|\eta_-|) < 1/\sqrt{3}$, 
$\bar \eta = \frac{1}{2}(\eta_0 + 1/\sqrt{3})$, $\bar\mu = \mu_-(\bar\eta)$,
$\bar a = a(\bar \eta) = (1-3\bar \eta^2)/(1-\bar \eta^2)$, and we assume 
that $\epsilon \le \bar \eta - \eta_0$. This implies in particular that
$|\eta(\tau,\xi)| = |\n(\tau,\xi) + \delta(\tau,\xi)| \le \eta_0 + \epsilon 
\le \bar \eta < 1/\sqrt{3}$ for all $\xi \in \R$, $\tau \in [\tau_1,\tau_2]$. 

\blem Under\labell{lem6.1} the assumptions above, we have $E_0 \in 
\CC^1([\tau_1,\tau_2])$ and, for all $\nu > 0$, there exists a constant 
$K_2 > 0$ such that 
\beql{e4.11}
  \dot E_0(\tau) \,\le\, -\left(\frac{3}{2}-\nu\right)E_0(\tau) - 
  \bar a \int \delta^2\,\d\xi + K_2 \int {s'}^2\,\d\xi~,
\eeq
for all $\tau \in [\tau_1,\tau_2]$, where $\bar a = a(\bar \eta) = 
(1-3\bar \eta^2)/(1-\bar \eta^2) > 0$. 
\elem

\brem
Here and in the sequel, $K_2, K_3, \dots$ denote positive constants depending
on $\eta_+, \eta_-$, but not on the interval $[\tau_1,\tau_2] \subset \R_+$
nor on the solution $(\psi,s)$, provided $\|\psi(\tau)\|_{\H^2} + 
\|s(\tau)\|_{\H^1} \le \epsilon$. 
\erem
\PPP
By hypothesis on $\psi$, we have $E_0 \in \CC([\tau_1,\tau_2]$. To prove 
the differentiability, we first assume that the initial data $(\psi_1,s_1)$
belong to $\S(\R)$, the space of rapidly decreasing $\CC^\infty$ functions on 
$\R$. Then $\psi(\tau),s(\tau) \in \S(\R)$ for all $\tau \in [\tau_1,\tau_2]$,
$E_0 \in \CC^\infty([\tau_1,\tau_2])$, and a direct calculation shows that 
\beql{e4.12}
  \dot E_0(\tau) \,=\, - \int \psi' (A(\n + \psi') - A(\n))\,\d\xi - 
  \frac{3}{4}\int \psi^2\,\d\xi + \int \eta s' \psi\,\d\xi~.
\eeq
In the general case, we use the fact that the solution $(\psi(\tau),s(\tau))$
depends continuously on the data $(\psi_1,s_1) \in \H^2(\R) \ti \H^1(\R)$, 
see Proposition~\ref{prop5.1}. If $F(\tau)$ denotes the right-hand side of 
\reff{e4.12}, we see that (for fixed $\tau$) both $E_0(\tau)$ and 
$E_0(\tau_1) + \int_{\tau_1}^\tau F(t)\, \d t$ are continuous functions of
$(\psi_1,s_1)$. Since these functions coincide on a dense subset (namely, 
$\S^2(\R)$), they must be equal everywhere. This shows that $E_0 \in 
\CC^1([\tau_1,\tau_2])$ and satisfies \reff{e4.12}. 

To prove \reff{e4.11}, we first note that $A(\n + \psi')-A(\n) = a(\hat \eta)
\psi'$ for some $\hat \eta \in [\n, \n+\psi']$. Since $|\hat \eta| \le \bar 
\eta$ by assumption, we have $\psi' (A(\n + \psi')-A(\n)) = a(\hat \eta)
{\psi'}^2 \ge \bar a {\psi'}^2$. On the other hand, for all $\nu > 0$, 
one has $|\eta s' \psi| \le |s' \psi| \le \frac{\nu}{2} \psi^2 + \frac{1}
{2\nu} {s'}^2$. We thus find
\[
  \dot E_0(\tau) \,\le\, - \bar a \int {\psi'}^2\,\d\xi - 
  \left(\frac{3}{4} - \frac{\nu}{2}\right) \int \psi^2 \,\d\xi + 
  K_2 \int {s'}^2\,\d\xi~,
\]
where $K_2 = (2\nu)^{-1}$. This concludes the proof of Lemma~\ref{lem6.1}. 
\QQQ

\blem Under\labell{lem6.2} the same assumptions, we have $E_1 \in 
\CC^1([\tau_1, \tau_2])$, and there exists a constant $K_3 > 0$ such 
that 
\[
  \dot E_1(\tau) \,\le\, -\bar \mu E_2(\tau) - \frac{1}{2} \e^\tau \int s^2 
  \,\d\xi + K_3 \left(\e^{-\tau} + \int(s^2 + \delta^2)\,\d\xi\right)~,
\]
for all $\tau \in [\tau_1,\tau_2]$, where $\bar \mu = \mu_-(\bar \eta)$ is 
given by \reff{e4.10}.  
\elem
\PPP
As in the proof of Lemma~\ref{lem6.1}, the differentiability of $\int \delta^2
\d\xi$ and $\int s^2 \d\xi$ is easily verified using a density argument. From
\reff{e2.13}, we have
\[\ba{rcl}
  \int \delta \dot \delta \,\d\xi &=& - \int \delta' (A(\n+\delta)-A(\n))'
   \,\d\xi -\frac{1}{4} \int \delta^2\,\d\xi - \int \eta s' \delta' \,\d\xi \\
  &=& - \int(a(\eta){\delta'}^2+\eta s'\delta')\,\d\xi - \frac{1}{4} \int 
   \delta^2\,\d\xi - \int \n'\delta'(a(\n+\delta)-a(\n))\,\d\xi~.
\ea\]
Since $|\n'\delta'(a(\n+\delta)-a(\n))| \le |\n'||a'(\bar \eta)| |\delta 
\delta'| \le \frac{\bar \mu}{4}{\delta'}^2 + C_1 \delta^2$ for some $C_1 > 0$,
we obtain
\beql{e4.13}
  \int \delta \dot \delta \,\d\xi \,\le\, - \int(a(\eta){\delta'}^2+\eta s'
  \delta')\,\d\xi + \frac{\bar \mu}{4} \int {\delta'}^2\,\d\xi + 
  C_1 \int \delta^2 \,\d\xi~.
\eeq

On the other hand, using \reff{e2.12} and integrating by parts, we find
\[\ba{rcl}
  \int s \dot s\,\,\d\xi &=& -\int(a_1(\eta){s'}^2+a_2(\eta)s'\delta')\,\d\xi
   - \frac{1}{4} \int s^2\,\d\xi - 2\e^\tau \int(1-\eta^2)s(\e^s-1)\,\d\xi \\
  && +\int \left(\frac{1}{2}s{s'}^2 - a_1'(\eta)\eta'ss' + a_2(\eta)\n''s
   -a_2'(\eta)\eta'\delta's + a_3(\eta)s {\eta'}^2\right)\,\d\xi~.
\ea\]
Since $\eta^2 \le 1/3$ and $|s| \le \epsilon \le 1/2$, we have $2(1-\eta^2)
s(\e^s-1) \ge s^2$. The other terms are bounded as follows: $|s{s'}^2| \le
\epsilon {s'}^2$, $|a_1'(\eta)\eta'ss'| \le C|(\n'+\delta')ss'| \le C(|ss'|+
\epsilon|s'\delta'|)$, $|a_2(\eta)\n''s| \le C|s|\e^{-|\xi|}$, $|a_2'(\eta)
\eta'\delta's| \le C(|s\delta'|+\epsilon{\delta'}^2)$, $|a_3(\eta)s{\eta'}^2|
\le C(|s|\e^{-|\xi|}+\epsilon{\delta'}^2)$. We thus obtain
\[\ba{rcl}
  \int s \dot s \,\d\xi &\le& -\int(a_1(\eta){s'}^2+a_2(\eta)s'\delta')\,\d\xi
   -\left(\frac{1}{4} + \e^\tau\right)\int s^2 \,\d\xi \\
  && + C_2 \int (|s|\e^{-|\xi|} + |ss'| + |s\delta'| + \epsilon({s'}^2 + 
   {\delta'}^2))\,\d\xi~,
\ea\]
for some $C_2 > 0$. Now, assuming that $\epsilon \leq \bar\mu/(4C_2)$, 
we have $C_2 \epsilon ({s'}^2 {+} {\delta'}^2) \le \bar \mu ({s'}^2 {+} 
{\delta'}^2)/4$. In addition, $C_2 |s|\e^{-|\xi|} \le \frac{1}{2} s^2 \e^\tau 
+ C \e^{-\tau} \e^{-2|\xi|}$ and $C_2(|ss'|+|s\delta'|) \le \bar \mu ({s'}^2+
{\delta'}^2)/4 + C s^2$. Therefore, 
\beql{e4.14}\ba{rcl}
  \int s \dot s \,\d\xi &=& -\int(a_1(\eta){s'}^2+a_2(\eta)s'\delta')\,\d\xi
   + \left(C_3 - \frac{1}{2}\e^\tau\right) \int s^2 \,\d\xi \\
  &&+ \frac{\bar \mu}{2}\int ({s'}^2 + {\delta'}^2)\,\d\xi + C_3 \e^{-\tau}~, 
\ea\eeq
for some $C_3 > 0$. 

Finally, by definition of $\mu_-(\eta)$ (see \reff{e4.10}), we have the 
inequality
\beql{e4.15}
  \int\left(3a(\eta){\delta'}^2 + (3\eta+a_2(\eta))s'\delta' + a_1(\eta)
  {s'}^2\right)\,\d\xi \,\ge\, 2\mu_-(\bar \eta)E_2 \,=\, 2 \bar \mu E_2~.
\eeq
Thus, multiplying \reff{e4.13} by $3$ and adding the result to \reff{e4.14}, 
we obtain
\[
  \dot E_1(\tau) \,\le\, - \bar \mu E_2(\tau) + \left(C_3 - \frac{1}{2}
  \e^\tau\right)\int s^2 \,\d\xi + 3C_1 \int \delta^2 \,\d\xi + C_3 
  \e^{-\tau}~,
\]
for all $\tau \in [\tau_1,\tau_2]$. This concludes the proof of 
Lemma~\ref{lem6.2}. 
\QQQ

\blem 
Under\labell{lem6.3} 
the same assumptions, $(\psi,s) \in
\L^2([\tau_1,\tau_2],\H^3(\R)\ti\H^2(\R))$, $E_2 \in 
\W^{1,1}([\tau_1, \tau_2])$, and there exists a constant $K_4 > 0$ such 
that 
\beql{e4.16}
  \dot E_2(\tau) \le -\bar \mu E_3(\tau) - \frac{1}{2} \e^\tau \int {s'}^2 
  \,\d\xi + K_4 \left(\e^{-\tau} + \e^\tau \int s^2 \,\d\xi + \int({s'}^2 + 
  \delta^2 + {\delta'}^2)\,\d\xi\right)~.
\eeq
\elem

\brem Due to the regularization, one also has $(\psi,s) \in \CC((\tau_1,
\tau_2],\H^3(\R)\ti\H^2(\R))$, $E_2 \in \CC^1((\tau_1,\tau_2])$, and 
\reff{e4.16} holds for all $\tau \in (\tau_1,\tau_2]$. 
\erem
\PPP
As in the proof of Lemma~\ref{lem6.1}, we first assume that $\psi(\tau), 
s(\tau) \in \S(\R)$ for all $\tau \in [\tau_1,\tau_2]$. Then, we have from 
\reff{e2.13}
\beql{e4.17}
  \int \delta'\dot\delta'\,\d\xi \,=\, - \int \delta''\dot\delta \,\d\xi
  \,=\, - \int \delta'' (A(\n+\delta)-A(\n))'' \,\d\xi + \frac{1}{4} 
  \int {\delta'}^2\,\d\xi - \int \delta'' (\eta s')'\,\d\xi~.
\eeq
Since $(A(\n+\delta)-A(\n))'' = a(\eta)\delta'' + (a(\n+\delta) - a(\n))\n''
+ (a'(\n+\delta)-a'(\n))(\n')^2 + a'(\eta)(2\n'+\delta')\delta'$ and $(\eta
s')' = \eta s'' + (\n' + \delta')s'$, we obtain using straightforward bounds
\[\ba{rcl}
  \int \delta'\dot\delta'\,\d\xi &\le& -\int (a(\eta){\delta''}^2 + 
   \eta s'' \delta'')\,\d\xi + \frac{1}{4}\int {\delta'}^2\,\d\xi \\
  && + C_4 \int |\delta''|(|\delta| + |\delta'| + |s'| + |{\delta'}^2| 
   + |s' \delta'|)\,\d\xi~,
\ea\]
for some $C_4 > 0$. Now, $C_4 |\delta''|(|\delta|+|\delta'|+|s'|) \le 
\frac{\bar \mu}{8}{\delta''}^2 + C(\delta^2 + {\delta'}^2 + {s'}^2)$ and
\[\ba{rcl}
  C_4 \|\delta'' \delta' (|\delta'|+|s'|)\|_{\L^1} &\le& 
   C_4 \|\delta''\|_{\L^2}^{3/2}\|\delta'\|_{\L^2}^{1/2}(\|\delta'\|_{\L^2}
   + \|s'\|_{\L^2}) \,\le\, 2 \epsilon C_4 \|\delta''\|_{\L^2}^{3/2}
   \|\delta'\|_  {\L^2}^{1/2} \\
  &\le& \frac{\bar \mu}{8} \|\delta''\|_{\L^2}^2 + C\epsilon^4 
  \|\delta'\|_{\L^2}^2~,
\ea\]
hence
\beql{e4.18}
  \int \delta'\dot\delta'\,\d\xi \,\le\, -\int (a(\eta){\delta''}^2 + 
   \eta s'' \delta'')\,\d\xi + \frac{\bar \mu}{4}\int {\delta''}^2\,\d\xi\\
  + C_5 \int (\delta^2 + {\delta'}^2 + {s'}^2)\,\d\xi~,
\eeq
for some $C_5 > 0$. 

On the other hand, assuming always $\psi(\tau),s(\tau) \in \S(\R)$, we have 
from \reff{e2.12}
\beql{e4.19}\ba{rcl}
  \int s' \dot s' \,\d\xi &=& - \int s'' \dot s \,\d\xi \,=\, 
   -\int (a_1(\eta){s''}^2 + a_2(\eta) s'' \eta'')\,\d\xi \\
  && + \frac{1}{4} \int{s'}^2\,\d\xi + 2\e^\tau \int s''(1-\eta^2)(\e^s-1)
   \,\d\xi -\int a_3(\eta)s'' {\eta'}^2 \,\d\xi~.
\ea\eeq
Integrating by parts, we obtain
\[
  -\int a_2(\eta)s''\n''\,\d\xi \,=\, \int s'(a_2(\eta)\n''' + a_2'(\eta)\n''
   (\n'+\delta'))\,\d\xi \,\le\, C\int |s'|(\e^{-|\xi|} + |\delta'|)\,\d\xi~,
\]
\[\ba{rcl}
  -\int a_3(\eta)s'' {\eta'}^2\,\d\xi &=& \int s'(a_3'(\eta){\eta'}^3 + 
   2a_3(\eta)\eta'\eta'')\,\d\xi \\
  &\le& C\int |s'|(\e^{-|\xi|}+|\delta'|+|\delta''|+|{\delta'}^3|+|\delta'
   \delta''|)\,\d\xi~,
\ea\]
\[\ba{rcl}
  2 \e^\tau \int s''(1-\eta^2)(\e^s -1)\,\d\xi &=& -2\e^\tau \int(1-\eta^2)
   \e^s {s'}^2 \,\d\xi + 4\e^\tau \int \eta\eta' (\e^s-1)s' \,\d\xi \\
  &\le& -\e^\tau \int {s'}^2\,\d\xi + C \e^\tau \int |ss'|(1+|\delta'|)\,
   \d\xi~.
\ea\]
Therefore, 
\[
  \int s'\dot s' \,\d\xi \,=\, -\int (a_1(\eta){s''}^2+a_2(\eta)s''\delta'')
  \,\d\xi + \left(\frac{1}{4} - \e^\tau\right) \int {s'}^2\,\d\xi + \int R
  \,\d\xi~,
\]
where $|R| \le C_6 |s'| (\e^{-|\xi|}+|\delta'|+|\delta''|+|{\delta'}^3|+
|\delta' \delta''|) + C_6 \e^\tau |ss'|(1+|\delta'|)$ for some $C_6 > 0$. 
Using the bounds $C_6 |s'|\e^{-|\xi|} \le \frac{1}{4}\e^\tau {s'}^2 + 
C \e^{-\tau} \e^{-2|\xi|}$ and
\[
  C_6 \|s'(|\delta'|+|\delta''|+|{\delta'}^3|+|\delta'\delta''|)\|_{\L^1}
  \,\le\, \frac{\bar \mu}{2}\|\delta''\|_{\L^2}^2 + C(\|s'\|_{\L^2}^2 +
  \|\delta'\|_{\L^2}^2)~,
\]
\[
  C_6 \e^\tau \|ss'(1+|\delta'|)\|_{\L^1} \,\le\, \frac{1}{4}\e^\tau 
  \|s'\|_{\L^2}^2 + C\e^\tau\|s\|_{\L^2}^2~,
\]
we find
\beql{e4.20}\ba{rcl}
  \int s'\dot s'\,\d\xi &\le& -\int(a_1(\eta){s''}^2 + a_2(\eta)s''\delta'')
   \,\d\xi + \left(C_7 - \frac{1}{2}\e^\tau\right)\int{s'}^2\,\d\xi \\
  && +\frac{\bar \mu}{2}\int{\delta''}^2\,\d\xi + C_7 \left(\e^{-\tau} +
   \int (\e^\tau s^2 + {\delta'}^2)\,\d\xi\right)~,
\ea\eeq
for some $C_7 > 0$. Finally, combining \reff{e4.18}, \reff{e4.20} and using 
the analogue of \reff{e4.15}, we arrive at \reff{e4.16}. This proves the claim
when $\psi(\tau), s(\tau) \in \S(\R)$. 

To infer the general case, we use again a density argument. Integrating
\reff{e4.16}, we obtain
\beql{e4.21}
  E_2(\tau_2) + \bar \mu \int_{\tau_1}^{\tau_2} E_3(\tau)\,\d\tau \,\le\, 
  E_2(\tau_1) + \int_{\tau_1}^{\tau_2} F(\tau)\,\d\tau~,
\eeq
where $F(\tau) \le C(\|\psi(\tau)\|_{\H^2}^2 + \|s(\tau)\|_{\H^1}^2)$ for 
all $\tau \in [\tau_1,\tau_2]$. Since $\S^2(\R)$ is dense in $\H^2(\R) \ti 
\H^1(\R)$, the estimate \reff{e4.21} shows that, if $(\psi,s) \in 
\CC([\tau_1,\tau_2], \H^2(\R) \ti \H^1(\R))$ is any solution satisfying the 
usual assumptions, then $(\psi,s) \in \L^2([\tau_1,\tau_2],\H^3(\R) \ti 
\H^2(\R))$. Using this result, it follows immediately from \reff{e4.17}, 
\reff{e4.19} that $\dot E_2 \in \L^1([\tau_1,\tau_2])$, hence $E_2 \in 
\W^{1,1}([\tau_1,\tau_2])$, and the same calculations as above show that 
\reff{e4.16} holds in the general case. This concludes the proof of 
Lemma~\ref{lem6.3}. 
\QQQ

Aa a consequence of Lemmas~\ref{lem6.1}, \ref{lem6.2}, \ref{lem6.3}, we have
the following estimate:

\bprop
There\labell{prop6.4} exist $\epsilon_2 > 0$, $K_5 \ge 1$ such that, 
if $(\psi,s) \in C([\tau_1,\tau_2],\H^2(\R) \ti \H^1(\R))$ is any solution 
of \reff{e2.12} satisfying $\|\psi(\tau) \|_{\H^2} + \|s(\tau)\|_{\H^1} \le 
\epsilon_2$ for all $\tau \in [\tau_1,\tau_2]$, then
\beql{e4.22}
  \|\psi(\tau)\|_{\H^2} + \|s(\tau)\|_{\H^1} \,\le\, K_5 \e^{-\tau/2} 
  \bigl(1+ \e^{\tau_1/2} (\|\psi(\tau_1)\|_{\H^2} + \|s(\tau_1)\|_{\H^1})
  \bigr)~,
\eeq
for all $\tau \in [\tau_1,\tau_2]$. 
\eprop
\PPP
Fix $\gamma \in (1/2,3/4)$ and let $\nu = 3/2-2\gamma$. We set $\epsilon_2 = 
\min(\bar \eta - \eta_0,\bar\mu/(4C_2))$, where $C_2$ is defined in the proof
of Lemma~\ref{lem6.2}. Then, for $A,B > 0$ large enough, we define
\[
  \Efour (\tau) \,=\, B(A E_0(\tau)+E_1(\tau)) + E_2(\tau)~, 
  \quad \tau \in [\tau_1,\tau_2]~.
\]
For instance, we may choose $\bar a A = K_3 + 11/4$ and assume that 
$B \ge 2K_4 +1$, $3 \bar \mu B \ge 2K_4 + 9/2$. Then, using 
Lemmas~\ref{lem6.1}, \ref{lem6.2}, \ref{lem6.3}, we find
\[
  \dot \Efour (\tau) \,\le\, -2\gamma \Efour (\tau) + \frac{1}{2}(C_8 - 
  \e^\tau)\int (s^2 + {s'}^2)\,\d\xi + C_8 \e^{-\tau}~, \quad \tau \in 
  [\tau_1,\tau_2]~,
\]
for some $C_8 > 0$. Integrating this inequality, we easily obtain
\[
  \Efour (\tau) \,\le\, C_9\left(\Efour (\tau_1)\e^{-2\gamma (\tau-\tau_1)}
  + \e^{-\tau}\right)~, \quad \tau \in [\tau_1,\tau_2]~,
\]
for some $C_9 > 0$. Since $2\gamma > 1$ and $\Efour (\tau)$ is equivalent 
to $\|\psi(\tau)\|_{\H^2}^2 +\|s(\tau)\|_{\H^1}^2$, this proves \reff{e4.22}.
\QQQ  

Combining Proposition~\ref{prop5.1} and Proposition~\ref{prop6.4}, we are now
able to complete the proof of Theorem~\ref{thm4.1}. 

\medskip\noindent{\bf Proof of Theorem~\ref{thm4.1}.}
Fix $\gamma \in (1/2,3/4)$, and let $\epsilon_3 =\min(\epsilon_1,\epsilon_2)$,
where $\epsilon_1$ is defined in Proposition~\ref{prop5.1} and $\epsilon_2$ 
in Proposition~\ref{prop6.4}. We set 
\[
   \epsilon_0 \,=\, \frac{\epsilon_3}{4K_5}~, \quad 
   \tau_0 \,=\, \max\left(0,-2\log\frac{\epsilon_3}{4K_5}\right)~,
\]
where $K_5$ is defined in Proposition~\ref{prop6.4}. Then, for all 
$(\psi_0,s_0) \in \H^2(\R) \ti \H^1(\R)$ satisfying $\|\psi_0\|_{\H^2} + 
\|s_0\|_{\H^1} \le \epsilon_0$, the system \reff{e2.12} has a unique 
global solution $(\psi,s) \in \CC([\tau_0,\infty),\H^2(\R) \ti \H^1(\R))$ 
with initial data $(\psi(\tau_0),s(\tau_0)) = (\psi_0,s_0)$. Indeed, 
according to our local existence result (Proposition~\ref{prop5.1}), 
it suffices to show that the solution $(\psi(\tau),s(\tau))$ 
satisfies $\|\psi(\tau)\|_{\H^2} + \|s(\tau)\|_{\H^1} < \epsilon_3$
whenever it exists. Assume on the contrary that there exists a time 
$\tau_3 > \tau_0$ such that $\|\psi(\tau_3)\|_{\H^2} + \|s(\tau_3)\|_{\H^1}
= \epsilon_3$ and $\|\psi(\tau)\|_{\H^2} + \|s(\tau)\|_{\H^1} < \epsilon_3$
for all $\tau \in [\tau_0,\tau_3)$. Then, according to
Proposition~\ref{prop6.4}, we have
\[
  \|\psi(\tau_3)\|_{\H^2} + \|s(\tau_3)\|_{\H^1} \,\le\, K_5 \e^{-\tau_3/2}
  \left(1+ \epsilon_0 \e^{\tau_0/2}\right) \,\le\, K_5\left(\e^{-\tau_0/2}
  +\epsilon_0\right) \,\le\, \frac{\epsilon_3}{2}~,
\]
which is a contradiction. Therefore, $(\psi(\tau),s(\tau))$ exists for all 
$\tau \ge \tau_0$ and satisfies $\|\psi(\tau)\|_{\H^2} + \|s(\tau)\|_{\H^1}
\le \epsilon_3/2$. In particular, we have from \reff{e4.22}
\beql{e4.24}
  \|\psi(\tau)\|_{\H^2} + \|s(\tau)\|_{\H^1} \,\le\, C \e^{-\tau/2}~,
  \quad \tau \ge \tau_0~,
\eeq
for some $C > 0$. 
It remains to show that $\|\psi(\tau)\|_{\H^2} + \|s(\tau)\|_{\H^1} = 
\OO(\e^{-\gamma \tau})$ as $\tau \to +\infty$. In fact, we shall prove that 
$\|\psi(\tau)\|_{\H^2} = \OO(\e^{-\gamma \tau})$ and $\|s(\tau)\|_{\H^1} = 
\OO(\e^{-\tau})$ as $\tau \to +\infty$.
 
We begin with $\|s(\tau)\|_{\L^2}$. From \reff{e4.14} and \reff{e4.24}, 
we have for all $\tau \geq \tau_0$:
\[
  \frac{d}{d\tau}\int s^2 \,\d\xi \le (\bar C_1 - \e^\tau) \int s^2\,\d\xi 
   + \bar C_1 \left(\e^{-\tau} + \int{\delta'}^2\,\d\xi\right)
  \le - \e^\tau \int s^2\,\d\xi + \bar C_2 \e^{-\tau}~,
\]
for some $\bar C_1, \bar C_2 > 0$. This differential inequality implies that 
$\|s(\tau)\|_{\L^2} = \OO(\e^{-\tau})$ as $\tau \to +\infty$. Indeed, let
$f(\tau) = \e^{2\tau}\|s(\tau)\|_{\L^2}^2$, and assume that $\e^\tau \geq 4$.
Then $\dot f(\tau) \leq 2f(\tau) + \e^{\tau}(\bar C_2 - f(\tau)) \leq e^{\tau}
(\bar C_2 - f(\tau)/2)$, hence $f(\tau) \leq 2\bar C_2 + 1$ if $\tau$ is 
sufficiently large.  

To prove that $\|s'(\tau)\|_{\L^2} = \OO(\e^{-\tau})$ as $\tau \to +\infty$,
we first have to show that $E_3(\tau) = \OO(\e^{-\tau})$, see \reff{e4.8}. 
This estimate can be established using exactly the same techniques as above. 
Indeed, due to the parabolic regularization, $E_3(\tau)$ is continuously 
differentiable for all $\tau > \tau_0$. Thus, proceeding as in 
Lemma~\ref{lem6.2} or \ref{lem6.3}, one verifies that there exists a $K_6 > 0$
such that 
\[\ba{rcl}
  \dot E_3(\tau) &\le& - \frac{1}{2} \e^{\tau} \int {s''}^2\,\d\xi + 
    K_6 \int ({s''}^2 + \delta^2 + {\delta'}^2 + {\delta''}^2)\,\d\xi \\
  && + K_6 \left(\e^{-\tau} + \e^\tau \int (s^2 + {s'}^2)\,\d\xi\right)~,
\ea\]
for all $\tau \ge \tau_0 + 1$ (say). Then, defining $\Efive (\tau) = 
D \Efour (\tau) + E_3(\tau)$ for some sufficiently large $D > 0$ and 
proceeding as in Proposition~\ref{prop6.4}, we obtain
\[
  \dot \Efive (\tau) \,\le\, -2\gamma \Efive (\tau) - \frac{1}{2}
  \e^{\tau} \int (s^2 + {s'}^2 + {s''}^2)\,\d\xi + K_7 \e^{-\tau}~,
\]
for some $K_7 > 0$. By Gronwall's lemma, we have
\beql{e4.25}
  \Efive (\tau) + \frac{1}{2}\int_{\tau_0+1}^{\tau} \e^{-2\gamma(\tau-t)}
  \e^t \|s(t)\|_{\H^2}^2\,\d t \,\le\, \bar C_3 \e^{-\tau}~, \quad 
  \tau \geq \tau_0+1~,
\eeq
for some $\bar C_3 > 0$. In particular, $E_3(\tau) = \OO(\e^{-\tau})$ as $\tau 
\to +\infty$. Using this result, we deduce from \reff{e4.20}
\[\ba{rcl}
  \frac{d}{d\tau} \int {s'}^2 \,\d\xi &\le& \left(\bar C_4 -\e^\tau\right) 
  \int {s'}^2\,\d\xi + \bar C_4 \left(\e^{-\tau} + \int (\e^\tau s^2 + 
  {\delta'}^2 + {\delta''}^2)\,\d\xi\right) \\
  &\le& \left(\bar C_4 -\e^\tau\right) \int{s'}^2\,\d\xi + \bar C_5 
  \e^{-\tau}~,
\ea\]
hence $\|s'(\tau)\|_{\L^2} = \OO(\e^{-\tau})$ as $\tau \to +\infty$. 

Finally, choosing $A, B > 0$ large enough, we define
\[
  \Esix (\tau) \,=\, B \left(A \int \psi^2\,\d\xi + \int \delta^2 \,\d\xi
  \right)+\int {\delta'}^2\,\d\xi~.
\]
Using \reff{e4.11}, \reff{e4.13}, \reff{e4.18} and proceeding as in the 
proof of Proposition~\ref{prop6.4}, we obtain
\beql{e4.26}
 \dot \Esix (\tau) \,\le\, -2\gamma \Esix (\tau) + \bar C_6  
 \|s(\tau)\|_{\H^2}^2~,
\eeq
for some $\bar C_6 > 0$. To control the last term in the right-hand side, we 
recall that
\[
  \int_{\tau_0+1}^\tau \e^{(2\gamma+1)t} \|s(t)\|_{\H^2}^2\,\d t
  \,\le\, 2\bar C_3 \e^{(2\gamma -1)\tau}~, \quad \tau \ge \tau_0+1~,
\]
by \reff{e4.25}. Since $2\gamma+1 > 2$, it follows that $\int_{\tau_0}^\infty
\e^{\lambda \tau} \|s(\tau)\|_{\H^2}^2\,\d\tau < \infty$ for all $\lambda 
< 2$. Therefore, we conclude from \reff{e4.26} that
\[
  \Esix (\tau) \le \e^{-2\gamma(\tau-\tau_0)} \Esix (\tau_0) + 
   \bar C_6 \int_{\tau_0}^\tau \e^{-2\gamma(\tau-t)} \|s(t)\|_{\H^2}^2\,\d t
  \le \bar C_7 \e^{-2\gamma \tau}~, 
\]
hence $\|\psi(\tau)\|_{\H^2} = \OO(\e^{-\gamma \tau})$ as $\tau \to +\infty$. 
The proof of Theorem~\ref{thm4.1} is now complete.  
\QQQ


\end{document}